\documentclass[12pt]{book}

\usepackage{microtype}

\usepackage[utf8]{inputenc}
\usepackage[T1]{fontenc}
\usepackage[backend=biber]{biblatex}
\usepackage[autoload=false, linenumbers=true, theme=default]{jlcode}
\usepackage{hyperref} 
\usepackage{tcolorbox} 
\usepackage{longtable} 
\usepackage{tocloft} 
\usepackage[export]{adjustbox} 
\usepackage{subcaption}
\usepackage{xurl} 
\usepackage{textcomp}
\usepackage{listings} 
\usepackage{pdflscape} 
\usepackage[a4paper,includehead,includefoot]{geometry}
\usepackage{import}
\usepackage{graphicx}
\usepackage{float}
\usepackage{multicol}
\usepackage{amsmath}
\usepackage{amsfonts}
\usepackage{hyperref}
\usepackage{fancyhdr}
\usepackage{amsthm}
\usepackage{mathtools}
\usepackage{algpseudocode}
\usepackage{algorithm}
\usepackage[section]{placeins}
\usepackage[colorinlistoftodos]{todonotes}

\theoremstyle{plain}

\theoremstyle{definition}
\newtheorem{defn}{Definition}[section]

\theoremstyle{remark}

\def\N{\mathbb{N}}

\DeclareUnicodeCharacter{0301}{*************************************}

\newcounter{review}[section]

\DeclarePairedDelimiter{\lrangle}{\langle}{\rangle}

\makeatletter
\providecommand{\subtitle}[1]{
  \apptocmd{\@title}{{\large #1}}{}{}
}
\makeatother

\title{Automated Code Optimization with E-Graphs\\}
\subtitle{Applied equality saturation in the presence of metaprogramming and multiple dispatch: generic and high-level expression rewriting and analysis in the Julia programming language.}
\author{Alessandro Cheli\\ \\ Bachelor Thesis in Computer Science, \\Tesi di Laurea in Informatica\\Università di Pisa\\ \\ Advisors:\\Prof. Gian-Luigi Ferrari\footnote{University of Pisa}\\Dr. Christopher Rackauckas\footnote{Massachussets Institute of Technology}\\Prof. Andrea Corradini\footnotemark[1]}

\date{December 3, 2021}

\newif\iffigures
\newif\iftables

\makeatletter
\let\OLDfigure\figure
\def\figure {\figures@in@document\OLDfigure }
\let\OLDtable\table
\def\table {\tables@in@document\OLDtable }
\let\OLDlongtable\longtable
\def\longtable {\longtables@in@document\OLDlongtable }

\def\figures@in@document {%
    \immediate\write\@mainaux {\global\string\figurestrue}%
    \global\let\figures@in@document\empty
}

\def\tables@in@document {%
    \immediate\write\@mainaux {\global\string\tablestrue}%
    \global\let\tables@in@document\empty
}

\def\longtables@in@document {%
    \immediate\write\@mainaux {\global\string\tablestrue}%
    \global\let\longtables@in@document\empty
}
\makeatother

\begin{document}

\maketitle
\tableofcontents
\iffigures
   \addcontentsline{toc}{section}{List of Figures}
   \listoffigures
\fi

\iftables
   \addcontentsline{toc}{section}{List of Tables}
   \listoftables
\fi


%
%

\chapter{Introduction}
\label{ch:intro}

This thesis proposes an advanced code and symbolic rewriting system for the Julia programming language. We show how it can practically solve some challenging problems, briefly outlined below.

\begin{enumerate}
    \item Can programmers implement their own high-level compiler optimizations for their domain-specific scientific programs, without the requirement of them being compiler experts at all?
    \item Can these optimizers be implemented by users in the same language they want to optimize? This is an instance of the so called \textit{two-language problem} in a new disguise! If so, can these optimizers be possibly embedded in the same programs that have to be optimized?
    \item Can these compiler optimizers be written in a high-level fashion without the need to worry about their \textit{ordering}?
    \item Can compiler optimizations in scientific domains be written \textit{as equations? As the mathematics that governs the science that we want to simulate on a machine?}.
    \item Can symbolic mathematics do high-level compiler optimizations? Or can compiler optimizers do high-level symbolic computation? What can go wrong if we make them the same thing? Is there even a programming language flexible enough to do this?
\end{enumerate}

This thesis provides preliminary answers to those questions. In particular, we will show that the Julia programming language is appropriate for developing a modern, high-level code rewriting and optimization framework that can break the barrier between symbolic mathematics and compiler optimizations. We will also discuss how our contributions benefit the ModelingToolkit.jl \cite{ma2021modelingtoolkit} modeling language. ModelingToolkit.jl is a modeling language implemented in Julia, designed for high-performance symbolic-numeric computation. It mixes ideas from symbolic computational algebra systems with causal and acausal equation-based modeling frameworks to give an extendable and parallel modeling system. It allows users to give a high-level symbolic description of a model in the form of DAEs (Differential Algebraic Equations), and uses symbolic manipulation algorithms to preprocess, analyze and enhance the models. Automatic transformations are applied before using numerical algorithms for solving, in order to make the system easily handle equations that could not be solved without symbolic intervention.

\section{The Two-Language Problem and Compiler Optimizations}

The two-language problem consists in users programming in a high-level language such as Python, R, MATLAB \cite{matlab2012matlab} or other alternatives, while the performance-critical parts have to be rewritten in other, lower-level languages such as C or C++ to counteract the poor performance of the high-level language. This is hugely inefficient, because it introduces a lot of wasted efforts, human error and most importantly it makes the design of such software systems overcomplicated, non-transparent, resulting in communication between developers being much more difficult than it should be. The two-language problem often arises in scientific computing, particularly when developing software that takes advantage of numerical methods. Many scientific software systems are often complex and comprised of many intercommunicating components that reside in different areas of computing.

A striking example of the two-language problem in scientific computing can be pointed out from the current state-of-the-art Python software ecosystem for deep learning. Deep learning encompasses knowledge from artificial intelligence, numerical computation, advanced calculus and programming. Thus, deep learning requires a great amount of complexity in co-operating software modules: Python is an interpreted language, and interpreters can cause a substantial performance loss over compiled languages. The core Python language was not designed for numerical computation, therefore users must rely on additional frameworks such as \textit{NumPy} in order to access numeric computation primitives and structures \cite{van2011numpy}. Frameworks, such as \textit{TensorFlow} \cite{abadi2016tensorflow}, implement the equivalent of completely new programming languages under the surface, accessible via complicated APIs that are responsible for a great amount of human error.

Going back to compiler optimizations, it is well known that they require extensive human knowledge to be developed, and often involve the usage of many, different programming languages that interact together, ending up in complicated instances of the \textit{two-language problem}.
Can we solve this problem by developing a system that allows programmers to implement customized compiler optimizations through high-level language constructs, with the goal of manipulating the same programming language in which the optimizations are described? Ideally, the users of such a system may want to write domain-specific optimization passes in the same programs they are developing. This hints that we need a language providing a powerful and structured metaprogramming abstraction, in order to write programs that can transform parts of themselves before being compiled. We want the users of our system to be able to achieve automatic program rewriting and optimization without being compiler experts, through a high-level syntax that resembles mathematics, and thus without ever having to dive into lower-level intermediate representations of code or compiler pipelines.


\section{Symbolic-Numerics and Compiler Optimizations}

\textit{Symbolic Computation}, also called Computer Algebra, focuses on the development of algorithms and software for manipulating syntactical (mostly mathematical) expression objects. Symbolic Computation emphasizes term rewriting (\cite{dershowitz1993taste}) over the actual numerical evaluation of code, whereas extracting a result from an expression implies the end of rewriteability. Term rewriting systems are reduction systems in which rewrite rules are used to transform symbolic expressions (terms). The manipulation of syntactical expression objects in computer science is not only relevant for mathematical applications; it is a fundamental component for the implementation, design and development of programming languages. For example, compilers themselves can be seen as many sequential passes of analysis and transformation of AST (Abstract Syntax Tree) objects \cite{aho1986compilers}. 

We can then define \textit{Symbolic-Numeric computation} as the use of algorithms in scientific computing that combine both symbolic and numeric methods to extend the domain of solvable problems. We refer to the book "Computer Algebra Handbook: Foundations, Applications, Systems" by Grabmeier, Johannes and Kaltofen \cite{grabmeier2003computer} for a broad survey on hybrid symbolic-numeric methods. The symbolic-numeric problems that we are going to discuss in this thesis are: 

\begin{enumerate}
    \item Symbolic pre-computation of expressions for efficient numerical evaluation.
    \item Code generation by computer algebra systems for solving numerical problems, for example PDE (Partial Differential Equation) solvers.
\end{enumerate}

Can a term rewriting system handle both program transformations in the host language and symbolic mathematics? Can compiler optimizations techniques be generalized to optimize symbolic-numeric computation?
An issue of many modern languages used for scientific computing is that they are not capable of representing programs written in them as structured data, and thus to modify their behavior before execution. This implies that many modern systems for computer algebra have been developed on a totally different abstraction level compared to the symbolic manipulation tools that handle executable code in the host language, even if the two tasks often involve expression manipulation through very similar term rewriting algorithms. An exception are languages of the Lisp family, where the extremely minimalistic syntax and computational model results in \textit{homoiconicity}, a property that consists in letting the primary representation of programs be a data structure in a primitive type of the language itself. This eases metaprogramming and allows for treating executable code with the same term rewriting algorithms that manipulate symbolic mathematics. We'll see how Julia provides many metaprogramming features similar to Lisp-like languages, particularly a code quoting and macro system very close to the ones available in the Scheme language. In popular folklore, many users even say that Julia is itself a dialect of Scheme. So why did we choose Julia?

\section{Why Julia}
\label{sec:whyjulia}


Julia is a recent programming language, first released in 2012, bringing a fresh approach to technical and numerical computing \cite{bezanson2017julia, bezanson2012julia}, disrupting the popular conviction that a programming language cannot be high-level, easy to learn, and performant at the same time. The Julia language is currently gaining a lot of traction in the field of technical computing: Julia’s innovation lies in its combination of productivity and performance, providing a pragmatical solution to the \textit{two-language problem} in the context of scientific and technical computing.

\begin{figure}
    \centering
    \includegraphics[width=.9\textwidth]{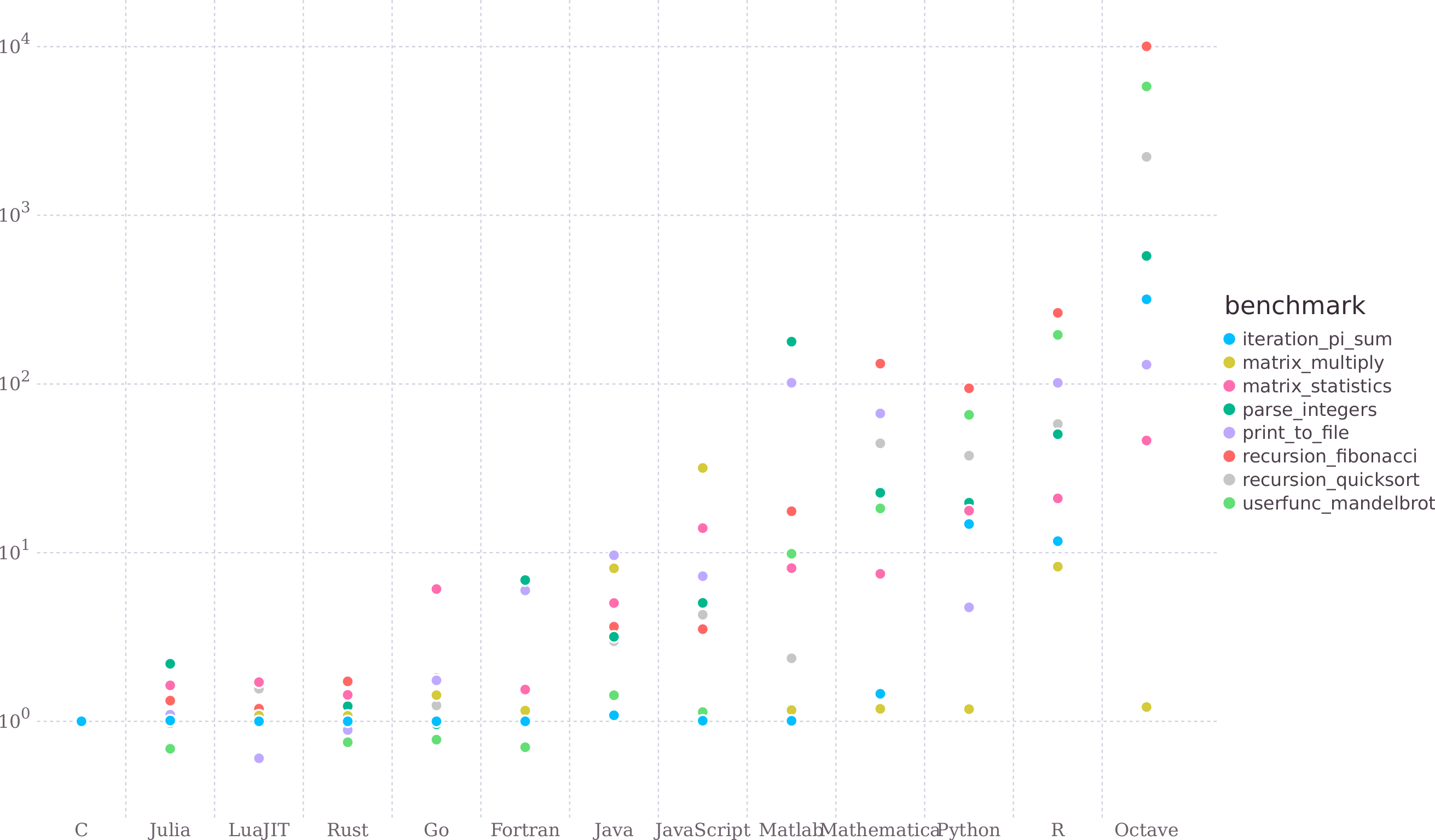}
    \caption{Micro-Benchmarks of some common algorithms in Julia}
    \label{fig:juliabenchmarks}
\end{figure}

In \autoref{fig:juliabenchmarks} we show some micro-benchmarks comparing Julia against other popular programming languages. The benchmarks are written to test the performance of identical algorithms and code patterns implemented in each language. The vertical axis shows each benchmark time against the C implementation. This figure is directly reported from the Julia website.\footnote{\href{https://julialang.org/benchmarks/}{https://julialang.org/benchmarks/}}

The most used languages for scientific computing are Python, R, MATLAB \cite{matlab2012matlab} and Mathematica \cite{wolfram1991mathematica}. Each has some weaknesses that Julia can hopefully solve: they are all part of the category of \textit{dynamically typed languages}: since those languages are easy to learn and write, many researchers today choose them for their technical computing research purposes. Applications written in interpreted and dynamically typed languages often result in a substantial performance loss when compared to programs written in compiled languages, thus, the aforementioned scientific computing languages resort to procedures written in lower-level languages to achieve speedups in the performance-critical sections of programs. This happens particularly for numerical procedures, and this is the critical instance of the two-language problem that we described above. 

The Julia language provides excellent numerical computing features. It provides an extensive library of mathematical functions with great numerical accuracy, making great use of vectorized code to gain a lot of performance benefits. There is primitive support for dense and sparse matrices, multithreading, distributed computing, SIMD instruction-level parallelism, an extensive hierarchy of number types, including complex numbers, n-dimensional arrays and an impressive set of linear algebra operations \cite{noack2015fast}.
The flexible and expressive type system gives Julia a boost over other scientific computing languages \cite{bezanson2017julia}. Julia allows for optional type annotations, allowing programmers to overload function names and operators: the language will automatically select the right code to execute based on the argument types. This paradigm called \textit{multiple dispatch} elegantly harmonizes with numerical computing, since many important mathematical operations involve interactions between many types of mathematical objects. It is intuitive to see that writing a program to compute the product of two matrices can be human friendly when using the same mathematical operator for computing scalar products, and that it should be the job of the compiler to automatically infer which low-level procedure is the most efficient to compute the operation.

Particularly adapt for symbolic computation is Julia's excellent metaprogramming and macro system, allowing for \textit{homoiconicity}: programmatic generation and manipulation of expressions as first-class values, a well-known paradigm found in Lisp dialects such as Scheme. (More details in \autoref{ssec:metaprogramming}). This feature has been fundamental for the development of the core contributions of this thesis. Being JIT-compiled, the emitted Julia code is compiled and executed in a fast runtime, relying on the LLVM compiler framework \cite{lattner2004llvm}. This allows for both fast debug cycles and for using Julia’s well-optimized numerical ecosystem with built-in parallelism in the generated code. 
The Julia core language and standard library also include an integrated package manager and high-performance parallelism and distributed computing functions, all without sacrificing the promise of being concise and easy to learn, a characteristic that also plays an important role in human-to-human communication.

To get an understanding of the role of the Julia programming language when compared to the myriad of available alternatives, we refer readers to the short article, "\textit{Why We Created Julia}"\footnote{\href{https://julialang.org/blog/2012/02/why-we-created-julia/}{https://julialang.org/blog/2012/02/why-we-created-julia/}} by Jeff Bezanson, Stefan Karpinski, Viral B. Shah and Alan Edelman, the designers of the Julia Programming language, published in February 2012.

\section{Structure and Organization of the Thesis}
The thesis is organized as follows:
\begin{itemize}
    \item \textbf{\autoref{ch:intro}} overviews some challenging problems in symbolic mathematics and compiler optimization, and why Julia is our programming language of choice for this framework.
    \item \textbf{\autoref{ch:solution}} introduces the package Metatheory.jl \cite{Cheli2021}, designed to provide generic term rewriting utilities for Julia code and symbolic expressions, and we are going to talk about its role in a redesign of the Julia symbolic mathematics package ecosystem. We then describe how Metatheory.jl provides a pure-Julia implementation of a novel technique called e-graph rewriting \cite{willsey2021egg} that generalizes an algorithm called equality saturation \cite{tate2009equality}, in order to provide a new abstraction over term rewriting that supports equational rewrite rules and is thus particularly suitable for high-level domain specific compiler optimizations and symbolic mathematics.
    \item \textbf{\autoref{ch:applications}} discusses a few real-world applications of the expression rewriting framework. We developed an example optimizer for functional streams in Julia, and we then discuss some results we have obtained when optimizing symbolic-numeric computations through e-graph rewriting, manipulating expressions from the Julia computer algebra system Symbolics.jl \cite{gowda2021high}.
    \item \textbf{\autoref{ch:future}} discusses future improvements, extensions and applications of our method. The e-graph rewriting technique and Julia's homoiconicity may allow us to go much further beyond compiler optimizations.
    \item We conclude the thesis with some final remarks, observations and acknowledgments in \textbf{\autoref{ch:conclusion}}.
\end{itemize}
\chapter{An Advanced Framework for Expression Rewriting in Julia}
\label{ch:solution}

In this chapter we will describe in detail the main contribution of this thesis: our proposed expression rewriting framework for the Julia programming language allows for automated symbolic expression optimization and analysis, relying on arbitrary equational theories that can be elegantly specified by users thanks to Julia's extensibility. Using Julia allowed us to solve the \textit{two-language} problem in the context of high-level program transformations: a goal of our proposed framework is to allow programmers to use those innovative symbolic manipulation techniques to develop specific compiler optimizations through symbolic code manipulation without ever leaving the comfort of the programming language in which they are writing their programs.
We will describe how a redesign of the structural architecture of packages in the Julia symbolic computation ecosystem resulted in better organization, feature decoupling and extended generality of this framework.
The key focus of this work resides in the innovative expression rewriting features offered by the Metatheory.jl \cite{Cheli2021} package, discussed in \autoref{sec:metatheory}, specifically term rewriting and code analysis through equality graphs, detailed in \autoref{ssec:egraphs}. Equality graphs \cite{willsey2021egg, nelson1980fast} (referred to as e-graphs from now on) are a data structure that elegantly extends the concept of union-find \cite{tarjan1975efficiency} data structures (also known as disjoint-sets)  to elegantly maintain the closure of a congruence relation over symbolic terms. This particular type of graphs has been used in the past in many automated theorem provers \cite{de2008z3}. Over the past decades, many projects have repurposed e-graphs to achieve excellent results in compiler optimization tasks based on source code rewriting through an algorithm called \textit{equality saturation} (\cite{tate2009equality, joshi02denali, panchekha2015automatically, nandi20szalinski, premtoon20yogo, stepp2011equality, wang20spores}). Many of these approaches, though, suffer from the two-language problem. We will see how e-graphs and equality saturation can be generalized to introduce a novel approach to general-purpose term rewriting \cite{willsey2021egg} that efficiently supports equational rules, infinite loops in terms and non-deterministic computations. We will also describe how this novel term-rewriting approach results in striking synergies with Julia's multiple dispatch, homoiconicity and extensible type system to effectively solve the two-language problem in the context of flexible symbolic manipulation and user-defined compiler optimization.
We will describe how, thanks to the excellent metaprogramming language features of Julia, e-graphs implemented in Metatheory.jl can be used to achieve automated program rewriting, allowing Julia programmers to introduce arbitrary compiler optimizations specifically tailored to their packages, without the requirement of them being compiler experts. All this can be achieved by end users through elegant and high-level reflective language features. In the next chapter we will then discuss the benefits of this optimization framework in practical applications. 

\section{A Layered Architecture for Julia Symbolics}
\label{sec:architecture}

Up until version 4, the Julia CAS (Computer Algebra System) Symbolics.jl \cite{gowda2021high} (\autoref{sec:symbolics}) was designed with a partially monolithic architecture. The underlying package for term rewriting, SymbolicUtils.jl, provided a generic interface for symbolic terms that allowed programmers to use the classical term rewriting utilities offered by the package on their own expression types. While our package Metatheory.jl (\autoref{sec:metatheory}) did exist at the time, it provided another similar interface of its own, not compatible with the Symbolics.jl ecosystem.

\begin{figure}
    \centering
    \includegraphics[width=\textwidth]{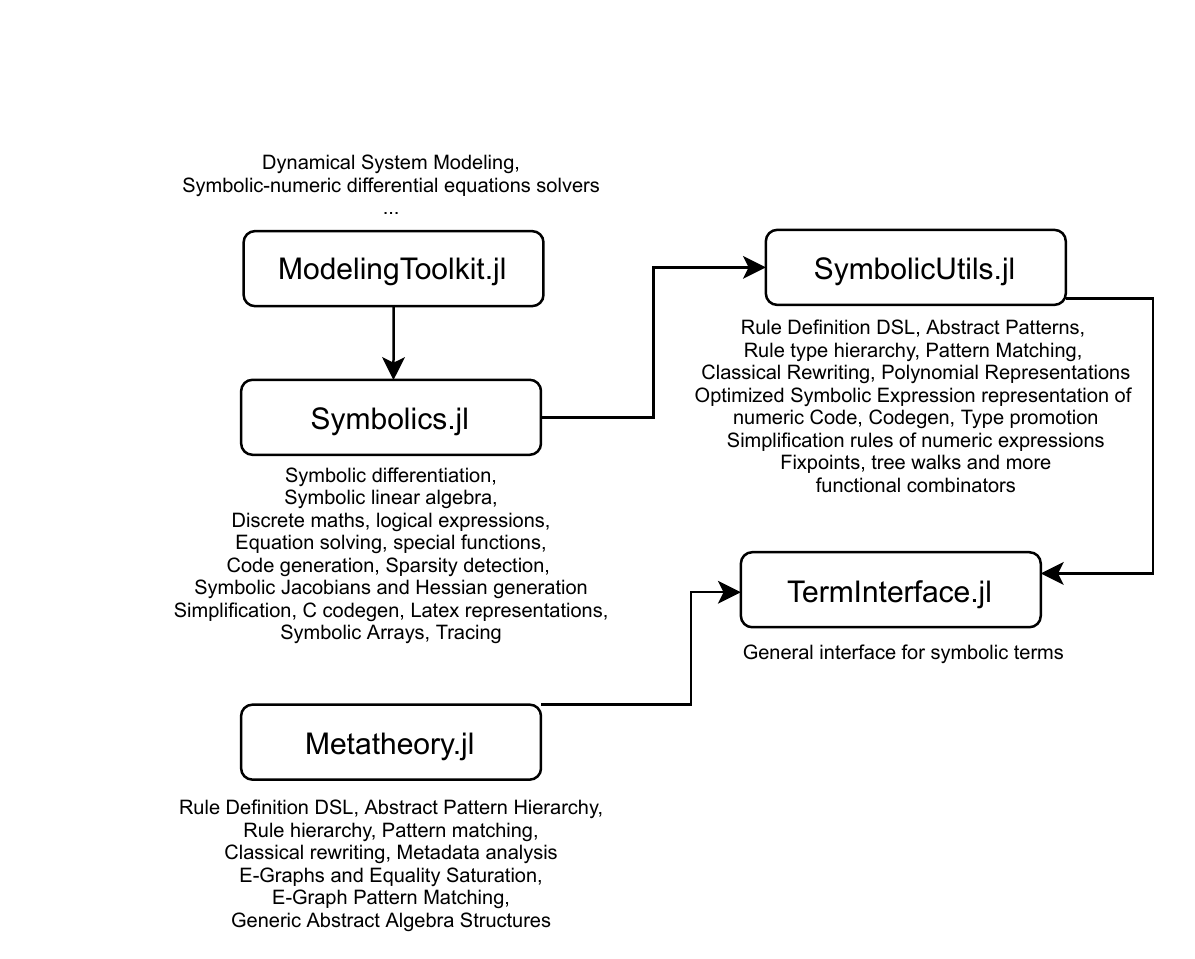}
    \caption{Simplified dependency graph illustrating the original state of the Julia symbolic-numeric ecosystem}
    \label{fig:symbolics-original-state}
\end{figure}

In their original state, many of the symbolic manipulation utilities that could have been otherwise be available as generic had been originally implemented with limited functionality in separated packages, working with specifically optimized expressions types or package-specific interfaces. Between Metatheory.jl and Symbolics.jl, this implied a lack of generality and a great amount of duplicated features, most importantly, two different eDSLs (embedded domain-specific languages) for rewrite rule definition and two different pattern matchers that shared the same term rewriting goals. Symbolics.jl performed well when manipulating specific mathematical expressions, while it was otherwise impossible to use the term rewriting features of Symbolics to rewrite homoiconic Julia expressions and therefore use the system to develop compiler optimizations through metaprogramming. In \autoref{fig:symbolics-original-state} we show a simplified diagram explaining the dependencies between some packages in the Julia symbolic computation ecosystem, and some of the relevant features they offered.

\begin{figure}
    \centering
    \includegraphics[width=\textwidth]{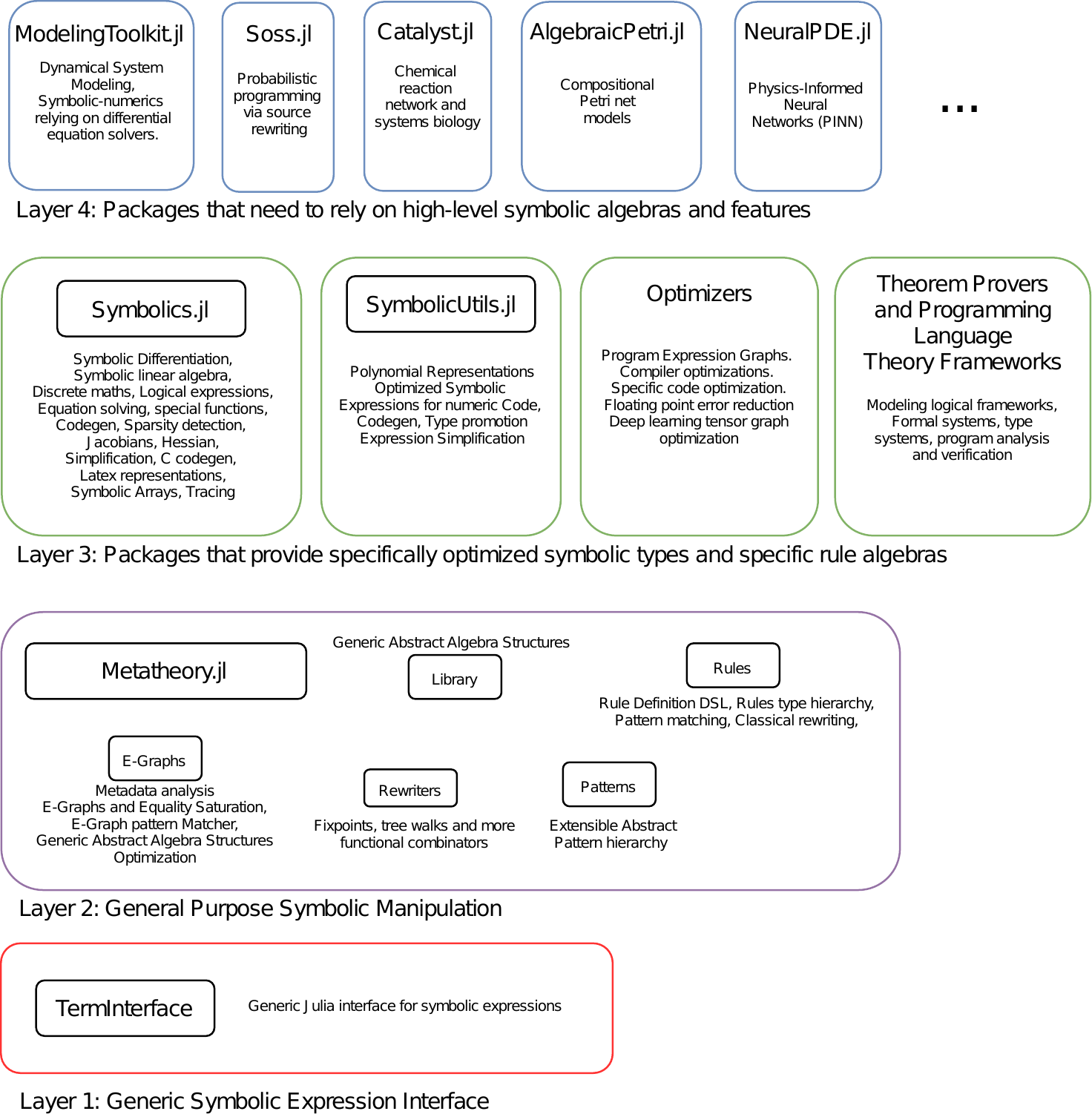}
    \caption{The novel layered architecture of the Julia symbolic computation ecosystem.}
    \label{fig:symbolics-layered-state}
\end{figure}

When integrating Symbolics.jl and Metatheory.jl for automated code optimization \cite{gowda2021high}, we decided to split the code defining the crucial interface for custom symbolic expression types into a different package called \textit{TermInterface.jl}, such that programmers defining their own symbolic expression types could implement methods to satisfy a single, shared interface, regardless of whether they want to use Metatheory.jl, Symbolics.jl or any other package that provides manipulation capabilities for symbolic expressions.  This drove us to design a novel, layered architecture for the ecosystem of Julia symbolic computation packages. In this chapter we will describe this novel architecture and how it can benefit the already existing Julia Symbolics stack by making Metatheory.jl's e-graph rewriting features available to all other packages in the higher layers, also allowing programmers to directly rewrite Julia code inside of Julia programs by using the same concepts and APIs. In this architecture, all the generalizable features have been decoupled from the packages where they have been introduced and have been merged into Metatheory.jl, a package that plays the role of a "lower level" building block for term rewriting. Packages in other layers of this architecture can then build more specific features and optimized symbolic representations on top of the generic layers. This architecture is illustrated in \autoref{fig:symbolics-layered-state}.

The advantages of a layered architecture are many. It reduces feature duplication and has allowed us to define a specification of the exposed programming interfaces, simplifying the work of documenting the various packages.
At the lowest layer, code in TermInterface.jl (\autoref{sec:terminterface}) is extremely generic, concise, and easy to learn. Packages that depend on it can follow the UNIX philosophy \cite{raymond2003art}, in short, they can do one thing and do it well.

In their original implementation, Metatheory.jl and SymbolicUtils.jl shared many duplicated features (such as classical rewriting utilities) and provided each their own, separate rule definition language and their own separate type hierarchy for rules and patterns. This clearly obstructed interoperability. By introducing the layered architecture, a single, extensible hierarchy of rewrite rules and patterns can exist for every package in this ecosystem. Thus, the rule definition eDSL can be shared from a lower layer, allowing packages in the higher layers to extend the rule definition language through generic abstract interfaces.
By having a single package expose both classical rewriting and e-graph rewriting features, packages on higher levels could use the two different term rewriting backends interchangeably; decoupling generalizable features from SymbolicUtils.jl and Symbolics.jl allows packages that require term rewriting features without a CAS, such as any potential theorem prover or compiler optimization package, to therefore avoid the dependency on Symbolics.jl or SymbolicUtils.jl's specialized symbolic mathematics code.

A layered architecture aids contributors in developing, testing and releasing the software in the ecosystem. 
Breaking changes between package versions can be easily handled, since a layered architecture aids the adaptation of semantic versioning \cite{dietrich2019dependency}, supported natively by Julia's package manager.
During software development, packages following this layered protocol can be tested against breaking changes through CI (Continuous Integration \cite{meyer2014continuous}) test suites that are incrementally run after every modification in the code is uploaded to the online code hosting service GitHub. When changes are uploaded for a specific package in the architecture, the service automatically triggers CI test suites, each running the core unit tests for other packages that depend on the one that is currently being developed.  If a dependent package test fails, the breakage is reported to the package maintainers and the newly introduced changes can be marked as breaking or fixed promptly.
By using semantic versioning and CI downstream testing, new non-breaking features introduced in lower layers are thus directly and safely available to higher layers without the need of developers having to release new versions of the dependent packages.

\section{TermInterface.jl}
\label{sec:terminterface}

In this section, we introduce the package \textit{TermInterface.jl} \footnote{Source code publicly available at \href{https://github.com/JuliaSymbolics/TermInterface.jl}{https://github.com/JuliaSymbolics/TermInterface.jl}}, providing the definition of an interface that any Julia symbolic expression type should satisfy. As the bottom layer of the architecture described in this work, this package plays a fundamental role for the rest of the packages in the symbolic computation ecosystem. 
During the development phase, we noticed how many term rewriting features should have been available as generic as possible, taking advantage of Julia's multiple dispatch to be able to handle any expression type satisfying the functions in an abstract interface. 
Releasing this package was the first step that inspired the overall redesign of the system into a layered architecture. Before this redesign, packages on higher levels of the architecture (like SymbolicUtils.jl) included many features such as an advanced classical term rewriting engine and a set of functional combinators, both designed specifically for highly specialized symbolic expression types provided in the package. Programmers were forced to rely on the specific type hierarchy for symbolic expressions provided by SymbolicUtils.jl. Introducing TermInterface.jl, allowed to decouple the actual term rewriting features from the underlying symbolic expression representation. Therefore, all the packages providing either symbolic manipulation utilities or specialized symbolic expression types should implement methods for the functions defined in TermInterface.jl to be able to use all the generic term rewriting features offered by Metatheory.jl and other packages offering symbolic manipulation capabilities.

\begin{defn}
    \textbf{TermInterface.jl Functions}

Given an expression type \texttt{T} and an atomic symbol type \texttt{S}, to satisfy \textit{TermInterface.jl} programmers must define the following methods. 

\begin{enumerate}
    \item \textbf{\texttt{istree(x::T)} and \texttt{istree(x::Type\{T\})}} \\
    Checks if \texttt{x} represents an expression tree, returning a boolean. If this function returns true, it is required that the \texttt{operation(::T)}, \texttt{exprhead(::T)} and \texttt{arguments(::T)}
    methods are defined.
    
    \item \textbf{\texttt{exprhead(x)}} \\ 
    If \texttt{x} is a term as defined by \texttt{istree(x)},
    \texttt{exprhead(x)} must return a symbol, corresponding to the head of
    the homoiconic Julia \texttt{Expr} most similar to the term \texttt{x}. If \texttt{x}
    represents a function call, for example, the \texttt{exprhead} should return the symbol
    \texttt{:call}. If \texttt{x} represents an indexing operation, such as
    \texttt{arr{[}i{]}}, then \texttt{exprhead} should return \texttt{:ref}. Note that
    \texttt{exprhead} is different from \texttt{operation} and both
    functions should be defined correctly in order to let other packages
    provide code generation and pattern matching features.

    \item \textbf{\texttt{operation(x::T)}} \\
    Returns the operation (a function object or a symbol) performed by an expression tree.
    This function should be called only if \texttt{istree(::T)} returns true. 
    
    \item \textbf{\texttt{arguments(x::T)}}\\
    Returns the arguments (a \texttt{Vector}) for an expression tree. Should be called
    only if \texttt{istree(x)} returns \texttt{true}. 

    \item \textbf{\texttt{similarterm(t::MyType,\ f,\ args,\ symtype=T;\ metadata=nothing,\\ exprhead=exprhead(t))}} \\
    and \\ \textbf{\texttt{similarterm(t::Type\{MyType\},\ f,\ args,\ symtype=T;\\ metadata=nothing,\ exprhead=:call)}.} \\
    This function should construct a new term with the operation \texttt{f} and arguments
    \texttt{args}, the term should be similar to \texttt{t} in type. For example, if
    \texttt{t} is a \texttt{SymbolicUtils.Term} object a new Term is created
    with the same \textit{symbolic type} as \texttt{t}. If not defined for a specialized expression type, the default result is computed as
    \texttt{f(args...)}. Defining this method for a custom term type will reduce
    any performance loss in performing \texttt{f(args...)}, especially on
    splatting, and redundant type computation. \texttt{T} is called the \textit{symbolic type} (\textit{symtype} in short) of the
    output term. The \texttt{exprhead} keyword argument is useful when converting 
    specific expression types to homoiconic Julia code, when \texttt{MyType == Expr}
\end{enumerate}

In addition, methods for \texttt{Base.hash} and
\texttt{Base.isequal} should also be implemented by the types for the
purposes of substitution and equality matching.

\end{defn}

\begin{defn}
\textbf{Optional functions in TermInterface.jl}

\begin{enumerate}
    \item \textbf{\texttt{unsorted\_arguments(x)}} \\ 
    If \texttt{x} is a term satisfying \texttt{istree(x)} and the term type
    \texttt{T} provides an optimized implementation for storing the
    arguments, this function can be used to retrieve the arguments when the
    order of arguments does not matter but the speed of the operation does.
    Defaults to \texttt{arguments(x)}.
    
    \item \textbf{\texttt{symtype(x)}} \\ 
The supposed type of values in the domain of \texttt{x}. Tracing tools can use
this type to pick the right method to run or analyse code.
This defaults to \texttt{typeof(x)}. 
\end{enumerate}
\end{defn}

\subsection{TermInterface.jl for homoiconic Julia expressions:}
\label{ssec:metaprogramming}
The Julia programming language provides excellent metaprogramming functionalities in its core standard library. Programmers can use values of type \texttt{Symbol} 
to represent identifiers in parsed Julia code (ASTs), or to represent a name or label to
identify an entity (e.g. as a dictionary key). Symbols can be entered using the \texttt{:} unary quote operator followed by a valid Julia identifier. The other type made available for representing homoiconic Julia programs is \texttt{Expr(head::Symbol, args...)}, representing compound expressions in parsed Julia code abstract syntax trees. Each expression consists of a head \texttt{Symbol}, identifying which kind of expression it is (e.g. a function call, a for loop, conditional statement, etc.), and subexpressions
(e.g. the name of the functions and the arguments of a call). The subexpressions are stored in a \texttt{Vector\{Any\}} field called \texttt{args}. Just like in the LISP dialect Scheme, programmers can use the quote operator \texttt{:} to create expression objects without using the explicit \texttt{Expr} constructor, and can later compute the result of quoted expression by calling the built-in \texttt{eval} function.
For example, entering \texttt{2+2} in a Julia REPL session, will display the result 4; entering the \textit{quoted expression} \texttt{:(2+2)} will return the unevaluated compound expression of type \texttt{Expr} to be later evaluated or manipulated. Julia provides a powerful macro system with macro hygiene that can be used to manipulate the contents of a program at compile time \footnote{More details on Julia metaprogramming can be found in the official documentation: \href{https://docs.julialang.org/en/v1/manual/metaprogramming/}{https://docs.julialang.org/en/v1/manual/metaprogramming/}}.  
Although this metaprogramming interface provided by default by Julia is very powerful, it is strictly low level. Julia does not provide any term rewriting or pattern matching system for homoiconic expressions in the standard library: to manipulate \texttt{Expr}s users must manually walk the AST and apply their manipulations. We will describe how such high-level expression manipulation features are available in the package Metatheory.jl in  \autoref{sec:metatheory}. TermInterface.jl provides a default implementation of its functions for Julia's built-in \texttt{Expr} type, such that homoiconic Julia code can be manipulated by any package exposing high-level symbolic manipulation, pattern matching and term rewriting functionalities.

\subsection{TermInterface.jl for specialized symbolic types:} In this example we are going to consider specialized expression types for symbolic mathematics, provided by SymbolicUtils.jl. For a term \texttt{t} of type \texttt{Add}, a Julia type providing an optimized representation for addition, \texttt{istree(t)} returns true, and
\texttt{operation(t)} returns the \texttt{+} generic function. \texttt{arguments(t)} returns
all the terms of the linear combination multiplied by the corresponding coefficient. Finally,
\texttt{symtype} represents the appropriate type, which is set when the term \texttt{t} is
created in Symbolics.jl, or is unused in other packages when not needed.
Term manipulation code can use \texttt{operation} and \texttt{arguments} after
checking if \texttt{istree} returns \texttt{true} on an object. Such code can
also use \texttt{similarterm} function to create a term. In fact, \texttt{Add} and
\texttt{Mul} were added to SymbolicUtils.jl system retroactively based on the above
interface, and the term manipulation functions continued to work as they
did.

\section{Metatheory.jl}
\label{sec:metatheory}


The main software contribution to this work consists in the Julia package
\textit{Metatheory.jl} \footnote{Source code publicly available at \url{https://github.com/JuliaSymbolics/Metatheory.jl}} \cite{Cheli2021}. 
Metatheory.jl is a general purpose term rewriting, metaprogramming and algebraic computation library for the Julia programming language, designed to take advantage of the powerful reflection capabilities to bridge the gap between symbolic mathematics, abstract interpretation, equational reasoning, optimization, composable compiler transforms, and advanced homoiconic pattern matching features. The core features of Metatheory.jl are a powerful rewrite rule definition language, a vast library of functional combinators for classical term rewriting and an e-graph rewriting system \cite{willsey2021egg}, a fresh approach to term rewriting achieved through equality saturation \cite{tate2009equality}. Metatheory.jl can manipulate any kind of Julia symbolic expression type, as long as it satisfies TermInterface.jl (\autoref{sec:terminterface}). One of the project goals of Metatheory.jl, beyond being easy to use and
composable, is to be fast and efficient.

Intuitively, Metatheory.jl transforms Julia symbolic
expressions into other symbolic expressions at both compile time and run time. This
allows users to perform customized and composable symbolic computation and compiler optimizations that are specifically tailored to arbitrary Julia programs. The library
provides a simple, algebraically composable interface to help scientists
implement and reason about all kinds of formal systems, by defining concise
rewriting rules as syntactically-valid Julia code. The primary benefit of using
Metatheory.jl is the algebraic nature of the specification of the rewriting
system. Composable blocks of rewrite rules bear a strong resemblance to
algebraic structures encountered in everyday scientific literature.

Metatheory.jl offers a concise eDSL (embedded Domain Specific Language), avaialble as plain Julia macros, allowing programmers to define \textit{rewrite rules} and \textit{theories}: composable
blocks of rewrite rules that can be executed through two, highly composable,
rewriting backends. The first is based on classical term rewriting. This
approach, however, suffers from the usual problems of rewriting systems. For
example, even trivial \textit{equational rules} such as commutativity may lead to
non-terminating systems and thus need to be adjusted by some sort of structuring
or rewriting order, which is known to require extensive user reasoning.

\begin{figure}
    \centering
    \includegraphics[width=.6\textwidth]{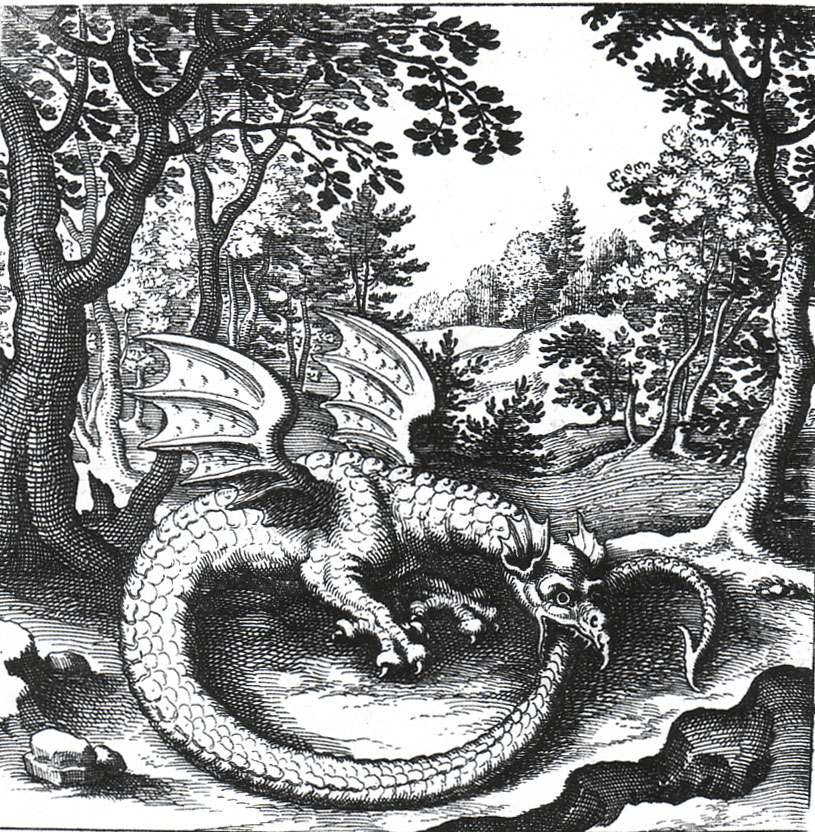} \\
    \vspace{5pt}
    \small{An engraving of a wyvern-type ouroboros by Lucas Jennis, in the 1625 alchemical tract De Lapide Philosophico.}
    \caption{The Metatheory.jl mascotte}
    \label{fig:dragon}
\end{figure}

The other back-end for Metatheory.jl, the core of our contribution, is designed
so that it does not require the user to reason about rewriting order. To do so
it relies on an a generalization of an algorithm called \textit{equality saturation} (\cite{tate2009equality, joshi02denali, panchekha2015automatically, nandi20szalinski, premtoon20yogo, stepp2011equality, wang20spores}), adoperating a data structure called  equality graph or \textit{e-graph} \cite{nelson1980fast}. The generalization of equality saturation for general purpose term rewriting was first introduced in the \textit{egg} Rust library \cite{willsey2021egg}.
\textit{E-graphs} can compactly represent many equivalent expressions and programs.
Provided with a collection of rewrite rules, defined in pure Julia, the \textit{equality
saturation} process iteratively executes an e-graph-specific pattern matcher and
inserts the matched substitutions. Since e-graphs can contain loops, infinite
derivations can be represented compactly and it is not required for the
described rewrite system be terminating or confluent. The main advantage of this approach 
is that users of the library can define \textit{bidirectional rewrite rules}, that can intuitively 
represent \textit{equational axioms} in any mathematical theory, without having to alter other rules in their rewrite system in order to maintain confluence.

The saturation process relies on the definition of e-graphs to include
\textit{rebuilding}, i.e. the automatic process of propagation and maintenance of
congruence closures. One of the core contributions of \texttt{egg} \cite{willsey2021egg} is a
delayed e-graph rebuilding process that is executed at the end of each
saturation step, whereas previous definitions of e-graphs in the literature
included rebuilding after every rewrite operation. Provided with \textit{equality
saturation}, users can efficiently derive (and analyze) all possible equivalent
expressions contained in an e-graph. The saturation process can be required to
stop prematurely as soon as chosen properties about the e-graph and its
expressions are proved. This latter back-end based on \textit{e-graphs} is suitable for
partial evaluators, symbolic mathematics, static analysis, theorem proving and
superoptimizers.

The original \texttt{egg} library \cite{willsey2021egg} is the first
implementation of generic and extensible term rewriting through equality saturation; the contributions of \texttt{egg} include novel
amortized algorithms for fast and efficient equivalence saturation and analysis.
Differently from the original Rust implementation of \texttt{egg}, which handles
expressions defined as Rust strings and data structures, our system directly
manipulates homoiconic Julia expressions and any expression type satisfying TermInterface.jl (\autoref{sec:terminterface}), and can therefore fully leverage
Julia's multiple dispatch and subtyping mechanisms \cite{zappa2018julia}, allowing programmers
to build expressions containing not only symbols but all kinds of Julia values.
This permits rewriting and analyses to be efficiently based on runtime data
contained in expressions. Most importantly, users can – and are encouraged to
– include predicate checks and type assertions in the left-hand side of rewrite rules.
Not only can Metatheory.jl manipulate the built-in Julia \texttt{Expr} type.
It is built on top of TermInterface.jl (explained in \autoref{sec:terminterface})
and can therefore rewrite on any type representing symbolic expressions that satisfy this
flexible interface. For example, integration between Symbolics.jl and Metatheory.jl
happens through this shared interface.

\begin{figure}
  \label{fig:metasyntax}
    \begin{align*}
      \text{function symbols}  & & \lrangle{fun}   & ::= & f \mid g \mid k \\ 
      \text{variables}          & & \lrangle{var}   & ::= & x \mid y \mid z \mid \hdots \\
      \text{terms (and patterns)}           &    & \lrangle{t}     & ::= &  \lrangle{fun}(\lrangle{t}, \hdots, \lrangle{t}) \mid \lrangle{var} \mid \lrangle{fun}  \\
        \text{ground terms}           &    & \lrangle{gt}     & ::= &  \lrangle{fun}(\lrangle{gt}, \hdots, \lrangle{gt}) \mid \lrangle{fun}  \\
    \text{e-class ids}          & & \lrangle{id}   & ::= & i \mid j  \\
      \text{e-nodes}       & & \lrangle{n}    & ::= & \lrangle{fun} \mid \lrangle{fun} (\lrangle{id}, \hdots, \lrangle{id})  \\
      \text{e-classes}          &  & \lrangle{c}    & ::= & \{ \lrangle{n}, \hdots, \lrangle{n}  \}  
    \end{align*}  
  \caption{Syntax and metavariables used for describing the Metatheory.jl system}
\end{figure}

\subsection{Rules and Patterns}
\label{ssec:patternsrules}

Two main components of the Metatheory.jl library are \textit{rewrite rules} and \textit{patterns}. Rewrite rules are composed of two patterns, namely the \textit{left hand side} and the \textit{right hand side} (\textit{LHS} and \textit{RHS} in short) and a \textit{rule operator}. Likewise, patterns are tree expressions, organized as Abstract Syntax Trees (ASTs), containing universally quantified variables used to \textit{match} against other tree-shaped expressions (ground terms) and to \textit{instantiate substitutions}. For readers familiar with functional programming languages, patterns in Metatheory.jl behave much like patterns in programming languages that have a primitive pattern matching system. To understand patterns and rules in detail we must first introduce some definitions from mathematical logic.

\begin{defn}
  \label{defn:terms}
  \textbf{Algebraic Signature and Terms} \\
  Let $\Sigma$ be a set of function symbols, we define the function for associated arieties as $ar$. An \textit{algebraic signature} is the pair $(\Sigma, ar)$. Constants are defined as functions in $\Sigma$ such that their ariety is 0. Let $V$ be a set of variables, then $T(\Sigma, V)$ is the \textit{set of terms} constructed from $\Sigma$ and $V$. The set of terms $T(\Sigma, V)$ is a set of syntactical expressions defined inductively as follows.
\end{defn}

\begin{enumerate}
\item
If \(x \in V\) then \(x \in T(\Sigma, V)\).
\item
\(\{k \in \Sigma \mid ar(k) = 0 \} \implies k \in T(\Sigma, V)\)
\item
If \(f \in \Sigma\) and \(ar(f) = n\) with \(n>0\) and \(t_1,\hdots,t_n\) are all contained in \(T(\Sigma, V)\) then \(f(t_1, \hdots, t_n) \in T(\Sigma, V)\). We call this form of terms \textit{\(f\)-applications}. If $f$ is a binary operator, then the syntax \(t_1 f t_2\) is also valid.
\end{enumerate}

\begin{defn}
\label{defn:ground_term}
  \textbf{Ground Terms} \\
  A ground term is a term in $T(\Sigma, V)$ that contains no variables, therefore, all terms in $T(\Sigma, \emptyset)$ are ground terms.
\end{defn}

We can now give an abstract definition of patterns and rewrite rules.

\begin{defn}
\label{defn:pattern}
  \textbf{Patterns} \\
  Non-ground terms in $T(\Sigma, V)$ are called \textit{patterns}.
\end{defn}

\begin{defn}
\label{defn:rules}
  \textbf{Rewrite Rules} \\
  Given two patterns $p_1,p_2 \in T(\Sigma, V)$, a \textit{(directed) rewrite rule} is written as $p_1 \rightarrow p_2$ to indicate that the left-hand side pattern $p_1$ can be replaced by the right-hand side pattern $p_2$. A (directed) term rewriting system is a set $R$ of such rules. A rule $p_1 \rightarrow p_2$ can be applied to a term $t$ if the left hand side pattern $p_1$ matches some subterm of $t$. More formally, if there is a substitution $\sigma$ such that the subterm of $t$ rooted at some position $pos$ is the result of applying the substitution $\sigma$ to the pattern $p_1$. The result term $t_1$ of this rule application is then the result of replacing the subterm at position $pos$ in $t$ by the pattern $p_1$ with the substitution $\sigma$ applied. Users can refer to \cite{dershowitz1993taste} for a more comprehensive survey on classical rewriting.
\end{defn}

\begin{defn}
\label{defn:patvars}
    \textbf{Pattern Variables} \\
    A variable $v \in V$ is called a pattern variable. When applying a rewrite rule $p_1 \rightarrow p_2$ to a term $t$, a pattern variable $v$ contained in $p_1$ can \textit{match against any subterm $t\prime$ of $t$}, just as long as the following occurences of $v$ in $p_1$ match against the same subterm $t\prime$ that has matched the first occurrence of the variable. This implies that in well-formed rules, every pattern variable that appears in the right-hand side pattern \textit{must be present in the left-hand side pattern.}  
\end{defn}

\begin{defn}
\label{defn:morerules}
    \textbf{Extensions of Patterns and Rewrite Rules} \\
    The concept of a Rewrite Rule can be easily extended. Bidirectional rewrite rules that can rewrite the LHS into the RHS and vice-versa are intuitively equivalent to equational axioms that permeate all about mathematics. It is not hard to see that introducing bidirectional rules in classical rewriting will cause computation loops and the rewrite systems will not terminate. The novel e-graph rewriting technique is used to circumvent this flaw. The concept of patterns can be easily extended too. In Metatheory.jl we provide a system and the corresponding high-level syntax to allow users to attach \textit{predicates} and \textit{type assertions} to pattern variables in rewrite rules, meaning that during pattern matching those variables will produce valid matches if and only if the attached predicates return true. This provides an elegant and efficient system to introduce \textit{conditional rewrite rules}. The Julia dynamic multiple dispatch mechanism makes it also easy to define two methods for the same predicate that will inspect both subterms during classical rewriting and non-deterministic equivalence classes during e-graph rewriting. 
    
\end{defn}

\subsubsection{Implementation of Rules in Metatheory.jl}
Metatheory.jl is not just limited to directed rewrite rules for symbolic manipulation. The package provides practical functionality for defining rewrite systems comprised of many different types of rules. Therefore, rules form a well-defined Julia type hierarchy. All rules are subtype of \texttt{AbstractRule}. The Julia programming language elegantly allows programmers to define special methods on user-defined types to treat them as \textit{callable functions}. Hence, Objects of type \texttt{<:AbstractRule} have been made callable and can be treated as first class function objects.
Directly calling a rule object with a single symbolic expression as argument will execute the pattern matcher in the context of classical rewriting, either returning the rewritten term if the argument expression matched the left-hand side pattern matched or returning the \texttt{nothing} value if a match was not produced. The presence of dynamic multiple dispatch in Julia allows us to define multiple methods when calling rule objects as functions, efficiently distinguishing the case when an e-graph object and an e-class id are passed as arguments to a rule, to then execute the specialized e-graph pattern matcher instead of the classical pattern matcher.

Metatheory.jl rules can be elegantly defined without explicitly calling the rule and pattern constructors using either the \texttt{@rule} or \texttt{@theory} macros, core of the Metatheory.jl eDSL. The \texttt{@rule} macro takes a pair of patterns, the left hand side and the right hand side and a binary operator called the \textit{rule operator} that defines which type of rule will be constructed by the eDSL. The \texttt{@theory} simply takes a block of Julia codes and calls \texttt{@rule} on each statement of the block, returning a flat vector of rules. 

The rule operators available in our system are:

\begin{enumerate}
    \item \texttt{LHS --> RHS}: creates a rule of type \texttt{RewriteRule}. The RHS is not evaluated but symbolically substituted on rewrite. This kind of rule behaves like regular directed rewrite rules.
    \item \texttt{LHS => RHS}: creates a rule of type \texttt{DynamicRule}. In this type of rules, the RHS is evaluated on rewrite. This is intuitively equivalent to defining an anonymous function guarded by pattern-matching.
    \item \texttt{LHS == RHS}: creates a rule of type \texttt{EqualityRule}. In e-graph rewriting, this rule behaves like \texttt{RewriteRule} but can go in both directions. Intuitively, this does not work in classical term rewriting.
   \item  \texttt{LHS $\ne$ RHS}: creates a rule of type \texttt{UnequalRule}. This kind of rule can only be used with the e-graphs backend, and is used to eagerly stop the process of rewriting if the \texttt{LHS} pattern is found to be equal to \texttt{RHS}.
\end{enumerate}

In \autoref{tbl:rules} we show what rule types are supported by the two rewriting backends.

\subsubsection{Implementation of Patterns in Metatheory.jl}

Just like ASTs, patterns in Metatheory.jl are organized as tree data structures. Nodes representing pattern variables and terms must be treated differently from other literal values. Thus, those special nodes are organized in a type hierarchy characterized by the abstract type \texttt{AbstractPat}, implementing methods from TermInterface.jl (\autoref{sec:terminterface}) with respect to the inductive definition given in \autoref{defn:terms}. Here we give a short outline of the \textit{pattern type hierachy}.

\begin{itemize}
    \item \textit{Pattern Variables}, also called \textit{slots}, are represented by the type \texttt{PatVar}, holding fields for the variable name, a number representing its order of appearance in the outermost parent pattern and optionally, a field containing a \textit{predicate} function that will be used to validate matches bt checking user-defined properties during pattern matching. Pattern variables can be written in the eDSL as a symbol prefixed with the tilde unary operator: \texttt{\~{}x}. Users can attach a predicate or a type check to a pattern variable by using the double colon operator followed by the name of the predicate function or the type to be checked \texttt{\~{}x::iszero} or \texttt{\~{}x::Int64} 
    \item \textit{Segment Patterns} represent a vector of subexpressions matched, and will match a variable number of subexpressions at once. Similarly to \texttt{PatVar}s, Segment Patterns are represented in the type \texttt{PatSegment} and can be written by prefixing the variable by a double tilde \texttt{\~{}\~{}x}, or analogously by using the \texttt{...} splatting operator: \texttt{\~{}x...}. Predicates can be attached to segment patterns and will be checked for every element in the vector of matched subexpressions. 
    \item \textit{Pattern Terms} are represented by the \texttt{PatTerm} type.
    Pattern terms will match on terms of the same arity and with the same function symbol and expression head, as defined in \autoref{sec:terminterface}.
    \item \textit{Literal Values} are not part of the \texttt{<:AbstractPat} hierarchy, such that any Julia literal value can be used as a pattern literal value to be matched against. Only requirement for matching against literals of a certain type is that methods for checking equality of two instances are well defined for that type. 
\end{itemize}

As an example we provide a definition of a simple symbolic rewrite rule, using the formula for the double angle of the sine function. As a recent feature, the \texttt{@rule} and \texttt{@theory} macros support adding the list of pattern variables names as the first arguments, in order to avoid using the \texttt{\~{}} operator. This makes definitions for rules and systems of rules much more readable and visually closer to textbook mathematics. 
\begin{jllisting}[language=julia, style=jlcodestyle]
using Metatheory

r1 = @rule x sin(2(x)) --> 2sin(x)*cos(x)
expr = :(sin(2z))
r1(expr)
# :((2 * sin(z)) * cos(z))
\end{jllisting}

\begin{table}
\centering
\caption{Comparison table of the different types of rewrite rules supported by Metatheory.jl.}
\label{tbl:rules}
\begin{tabular}{|l|l|l|l|l|} 
\hline
\textbf{Op.} & \textbf{Rule type} & \textbf{Description}                                                                                                                                                & \begin{tabular}[c]{@{}l@{}}\\\textbf{Classical}\\\textbf{Rewriting}\\\textbf{Support}\end{tabular} & \begin{tabular}[c]{@{}l@{}}\textbf{E-Graph}\\\textbf{Rewriting}\\\textbf{Support}\end{tabular}  \\ 
\hline
\texttt{-->}                & \texttt{RewriteRule}        & \begin{tabular}[c]{@{}l@{}}Symbolic rewrite rule \\that applies a symbolic \\substitution to the RHS.\end{tabular}                                                  & Yes                                                                                                     & Yes                                                                                                    \\ 
\hline
\texttt{=>}   & \texttt{DynamicRule}        & \begin{tabular}[c]{@{}l@{}}The RHS is evaluated \\as Julia code when \\applying the rule.\end{tabular}                                                              & Yes                                                                                                     & Yes                                                                                                    \\ 
\hline
\texttt{==}                & \texttt{EqualityRule}       & \begin{tabular}[c]{@{}l@{}}Behaves as a RewriteRule \\but~ bidirectionally. \\The RHS can also be \\rewritten into the LHS\end{tabular}                             & No                                                                                                      & Yes                                                                                                    \\ 
\hline
$\ne$, \texttt{!=}             & \texttt{UnequalRule}        & \begin{tabular}[c]{@{}l@{}}Used to detect \\contradictions in \\e-graph rewriting. \\If LHS is found equal\\ to RHS then halt the \\rewriting process.\end{tabular} & No                                                                                                      & Yes                                                                                                    \\
\hline
\end{tabular}
\end{table}

\subsubsection{Pattern Matcher for Classical Rewriting}

Before adapting the system to the layered architecture design, seen in \autoref{sec:architecture}, Metatheory.jl provided a rather naive classical pattern matcher while SymbolicUtils.jl provided a very flexible and performant implementation, although being strictly coupled to SymbolicUtils's own symbolic expression interface.  
With introduction of the novel Julia Symbolics ecosystem architecture,
Metatheory.jl's pattern matcher for classical term rewriting has been ported from SymbolicUtils.jl and has been extended to support a specialized pattern hierarchy and to match against any type satisfying TermInterface.jl.

The design of this pattern matcher is an adaptation of the excellent pattern matcher introduced in the book "\textit{Software Design for Flexibility}"  \cite{sdf} by Gerald Jay Sussman. Each rule object in Metatheory.jl stores a function object called the \textit{matcher}, responsible of matching terms against the pattern in the rule's left-hand side. Matchers are compiled from patterns at the time of rule construction through a procedure called the \textit{matcher compiler}. Matcher procedures are highly composable: each node in the pattern syntax tree represents a single matcher procedure. Since Julia is compiled, syntactically scoped and supports higher-order anonymous functions, as in many LISP dialects, the matcher compiler constructs the matcher procedures as regular Julia \textit{closures}, chained together by passing the next matcher as a callback function argument to the previous one. This results in a very performant pattern matcher optimized ahead of time by the Julia compiler.

\subsubsection{Rewriter Combinators}
\label{ssec:combinators}
Classical rewriting is useful for traversing the expression tree in a
specific way. In this regime, a rewriter is simply a function which
takes an expression and returns a modified expression.
Rules may be chained together into more sophisticated rewiters to avoid manual application of the rules. A rewriter is any callable object which takes an expression and returns another expression or the Julia value \texttt{nothing}. All rules in Metatheory.jl are therefore rewriters. If \texttt{nothing} is returned by a rule that means there were no changes applicable to the input expression.
Metatheory.jl contains a rewriter-combinator library for composing
multiple rules. The \texttt{Metatheory.Rewriters} module contains some types which create and transform rewriters. 

\begin{itemize}
    \item     \texttt{Empty()} is a rewriter which always returns \texttt{nothing}
    \item     \texttt{Chain(itr)} chains an iterator of rewriters into a single rewriter which applies each chained rewriter in the given order. If a rewriter returns \texttt{nothing} this is treated as a no-change.
    \item    \texttt{RestartedChain(itr)} behaves like the \texttt{Chain(itr)} rewriter but restarts from the first rewriter once on the first successful application of one of the chained rewriters.
    \item       \texttt{IfElse(cond, rw1, rw2)} runs the \texttt{cond} function on the input, applies \texttt{rw1} if \texttt{cond} returns \texttt{true}, \texttt{rw2} if it returns \texttt{false}
    \item      \texttt{If(cond, rw)} is the same as \texttt{IfElse(cond, rw, Empty())}
    \item     \texttt{Prewalk(rw; threaded=false, thread\_cutoff=100)} returns a rewriter which does a pre-order (from top to bottom and from left to right) traversal of a given expression and applies the rewriter \texttt{rw}. \texttt{threaded=true} will use multi threading for traversal. Note that if \texttt{rw} returns nothing when a match is not found, then \texttt{Prewalk(rw)} will also return \texttt{nothing} unless a match is found at every level of the walk. If an user is applying multiple rules, then the \texttt{Chain} rewriter already has the appropriate passthrough behavior. If users only want to apply one rule, then consider using \texttt{PassThrough}. \texttt{thread\_cutoff} is the minimum number of nodes in a subtree which should be walked in a threaded spawn.
    \item      \texttt{Postwalk(rw; threaded=false, thread\_cutoff=100)} similarly does post-order (from left to right and from bottom to top) traversal.
    \item         \texttt{Fixpoint(rw)} returns a rewriter which applies \texttt{rw} repeatedly until there are no changes to be made.
    \item      \texttt{FixpointNoCycle(rw)} behaves like Fixpoint but instead it applies rw repeatedly only while it is returning new results.
    \item \texttt{PassThrough(rw)} returns a rewriter which if \texttt{rw(x)} fails and returns \texttt{nothing} will instead return \texttt{x}, otherwise it will return \texttt{rw(x)}.

\end{itemize}

\subsection{E-Graphs and E-Graph Rewriting}
\label{ssec:egraphs}

In this section we describe the required background for understanding the process of \textit{E-Graph rewriting}.  
Classical rewriting falls short when the user's goal is to minimize a
certain cost function and the system of rewrite rules comprises many
rules interacting with one other. It is not straightforward to transform
an axiomatic formal system of equational rules into a Noetherian
(terminating) term-rewriting system, which is known to require a lot of
user reasoning \cite{dershowitz1993taste}. To circumvent these issues, there
has been newfound excitement in using equality saturation-based
rewriting engines for such requirements \cite{willsey2021egg}. This novel technique allows
users to define efficient term-rewriting systems with equational rules
without having to worry about termination or the ordering of rules,
resulting in algebraically compositional rewrite systems and efficient
algorithms for minimizing cost functions on chains of expression
rewrites. The Metatheory.jl \cite{Cheli2021} Julia package provides a
generic rewriting backend relying on the \textit{equality saturation} algorithm
and data structures called \textit{e-graphs} \cite{willsey2021egg}. 
For defining \textit{e-graphs}, associated concepts and procedures we adopt the naming conventions and syntax introduced in \cite{willsey2021egg} and
\cite{zhang2021relational}.

\begin{defn}
\label{defn:congruence_relation}
  \textbf{Congruence Relation} \\
  Given an equivalence relation $\equiv_\Sigma$, that is reflexive, transitive and symmetric over $T(\Sigma, \emptyset)$, we call $\cong_\Sigma$ the \textit{congruence relation} over $T(\Sigma, \emptyset)$ that satisfies the following implication, called the \textit{monotonicity axiom}. 
  We can omit the subscript $\Sigma$ when the context is clear

  $$ \forall f \in \Sigma . (\forall i \in \{1, \hdots, ar(f)\} . t_i \cong t_i^{\prime}) \implies f(t_1, \hdots, t_{ar(f)}) \cong f(t_1^{\prime}, \hdots, t_k^{\prime}) $$
\end{defn}


We have now come to the point of defining e-graphs.
\begin{defn}
    \label{defn:egraphintuitive}
  \textbf{E-Graph, intuitive definition} \\
  An Equality Graph \cite{nelson1980fast} (reffered to as e-graph in the rest of this work) is a data structure holding a set of symbolic terms and a congruence relation over the terms, composed of \textit{e-classes} and \textit{e-nodes}. \textit{E-classes} are sets of e-nodes that are considered equal, while \textit{e-nodes} consist of a function symbol $f$ and an ordered list of children \textit{e-class ids}.
  \textit{e-class ids} are opaque identifiers and can be simply represented with numbers in $\mathbb{N}$.
  Thus, an e-graph can be intuitively seen as a particular kind of \textit{bipartite graph} suited for symbolic computation with the notion of equivalence and congruence closure. Differently from tree data structures, e-graphs emphasize on sharing nodes for subterms.
\end{defn}

\begin{figure}[H]
    \centering
    \includegraphics[width=0.3\textwidth]{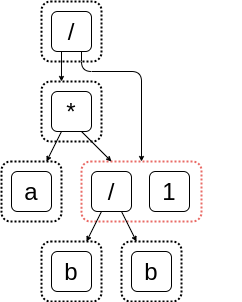}
    \caption{An example of an e-graph where $b/b$ is equivalent to 1.}
    \label{fig:egraph}
\end{figure}

In \autoref{fig:egraph} we see a simple egraph holding the symbolic term $\dfrac{a * (\frac{b}{b})}{1}$. The dashed boxes represent e-classes, while the regular boxes represent e-nodes. Given the simple algebraic equality $x/x = 1$, assuming that $x \ne 0$ for simplicity, the e-graph can compactly represent the two equivalent subterms $b/b$ and $1$ as the same entity by making them reside in the same \textit{e-class} (depicted in red).

\begin{defn}
    \label{defn:egraph}
  \textbf{E-Graph, formal definition} \\
  An \textit{e-graph} $G$ is a tuple $(U,M,lookup)$ where:
  \begin{enumerate}
      \item An \textit{union-find} data structure $U$, also known in literature as a \textit{disjoint sets} data structure, \cite{tarjan1975efficiency} storing the equivalence relation $\equiv_{id}$ over e-class ids. $U$ must provide a $find : (U, \mathbb{N}) \to  \N$ operation such that $find(U,i) = find(U, j) \iff i \equiv_{id} j$. Such operation \textit{canonicalizes} e-class ids, such that any e-class id $i$ is said to be canonical (over $U$) if and only if $find(U, i) = i$
      \item An \textit{e-class map} $M : \N \to \text{EClass}$ that maps e-class ids to e-classes (sets of e-nodes), such that all $\forall i,j \in U \mathbin{.} i \equiv_{id} j \iff M[i] \text{ is the same eclass } M[j]$.
      \item A function $lookup$, often implemented with an hashcons, that maps an e-node $n$ to the id $i$ of the e-class that contains it such that $n \in M[lookup(n)]$ where $lookup(n) = i$ and $i = find(i)$ ($i$ is canonical). 
      \item There are no duplicate e-nodes in the e-graph. An e-node with its head symbol and its children uniquely identifies the e-class that contains it.
  \end{enumerate}
\end{defn}


\subsection{Equality Saturation}
\label{ssec:eqsat}

Equality saturation (\cite{tate2009equality, joshi02denali, panchekha2015automatically, nandi20szalinski, premtoon20yogo, stepp2011equality, wang20spores}) is the procedure that consists in 1) creating an e-graph from an input program $p$, 2) iteratively rewriting the e-graph creating a vast set of equivalent programs equivalent to  $p$ and 3) extracting the "best" program from the e-graph according to an user-defined cost function.
An e-graph is said to be \textit{saturated} in relation to a set of rewrite rules if searching for all the rules and applying the relevant matches does not alter the structure of the e-graph. Intuitively, an e-graph saturates when all the possible rewrites have been applied and the e-graph already contains al the possible equivalent expressions. 

Compared to classical rewriting, adopting equality saturation lets programmers get rid of the tedious task of choosing when to apply which rewrites. The workflow is simpler: define the rewrite rules for the language in context, create an initial e-graph from a given expression, repeat the application rules until the e-graph is saturated or a limit of iterations is reached and finally extract the cheapest equivalent expression.

\begin{figure}
    \centering
    \includegraphics[width=.8\textwidth]{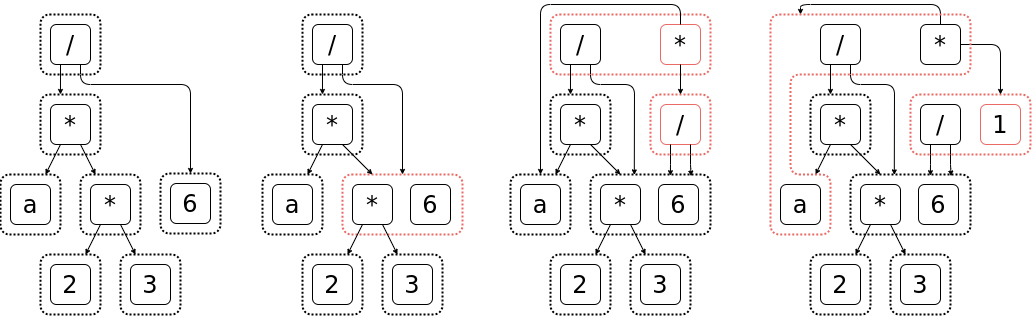}\\
    Equality saturation constructs the e-graph from a set of rules applied to an input expression. The four depicted e-graphs represent the process of equality saturation for the equivalent ways to write $a * (2 * 3) / 6$. The dashed boxes represent equivalence classes, and regular boxes represent e-nodes.
    \caption{Explanation of some equality saturation steps}
    \label{fig:eqsatpic}
\end{figure}

In \autoref{fig:eqsat} we give a definition of the algorithm in simplified pseudocode. The operations \textit{ADD} and \textit{MERGE} respectively add a new e-node to the e-graph and merge two e-classes in a single one. The \textit{LHS} and \textit{RHS} functions return the left-hand and right-hand patterns of a rewrite rule.
Even if our implementation supports bidirectional rewrite rules, for the sake of simplicity we do not describe how to achieve this behaviour in the  equality saturation pseudocode. The algorithm described in \autoref{fig:eqsat} can be extended to support bidirectional rules simply by converting them into two different oriented rules that go left-to-right and right-to-left. 
The \textit{EMATCH} function, short for \textit{e-graph pattern matching} searches an e-graph for all the possible matches corresponding to an arbitrary pattern. The fundamental difference from classical pattern matching is that a single pattern can yield a large amount of matches, due to the high degree of non-determinism in e-graphs. We will give a longer description of the e-matching procedure in \autoref{ssec:ematching}.
One of the novelties introduced in the generalized equality saturation procedure described in the egg paper \cite{willsey2021egg} is delayed rebuilding, that is restoring the congruence closure invariants at the end of each iteration. Delayed rebuilding walks an e-graph bottom-up and merges e-classes in order to maintain the congruence closure of the terms inside the e-graph (\autoref{defn:congruence_relation}) \cite{willsey2021egg}. In our simplified algorithm the \textit{REBUILD} function performs delayed rebuilding.
Readers can refer to the egg paper \cite{willsey2021egg} for more details about the equality saturation algorithm with delayed rebuilding. In \autoref{fig:eqsatpic} we instead show how applying some iterations of equality saturation affects an example e-graph, in the context of the example code in \autoref{fig:mtexample}.

\begin{figure}
    \centering
    \caption{Pseudocode for Equality Saturation}
    \label{fig:eqsat}
    \begin{algorithmic}[1]
        \Function{eqsat}{$expr, rules, timeout, costfun$}
            \State $g \gets \Call{egraph}{expr}$
            \State $i \gets 0$
            \While{$\neg \Call{saturated}{g} \wedge i \leq timeout$}
            \For{$r \in rules$}
            \State $lhs \gets \Call{lhs}{r}$
            \State $rhs \gets \Call{rhs}{r}$
            \State $matches \gets \Call{ematch}{g, lhs}$
            \For{$(sub, eclass) \in matches$}
                \State $newexpr \gets \Call{subst}{rhs, sub}$
                \State $neweclass \gets \Call{add}{g, newexpr}$
                \State $\Call{merge}{g, eclass, neweclass}$
            \EndFor
            \EndFor
            \State $\Call{rebuild}{g}$
            \State $i \gets i + 1$
            \EndWhile\\
            \Return $\Call{extract}{g, costfun}$ 
        \EndFunction
    \end{algorithmic}
\end{figure}

\subsubsection{Implementation Details of Equality Saturation}

In our implementation of equality saturation in Metatheory.jl, users can provide the algorithm a wide range of parameters, specified in the \texttt{SaturationParams} type (\autoref{fig:saturationparams}).
After constructing an e-graph from an expression and a set of rewrite rules using the \texttt{@theory} and \texttt{@rule} macros, users can optionally construct an object of type \texttt{SaturationParams} and execute equality saturation by calling the \texttt{saturate!(egraph, rules, parameters)} function. The \texttt{timeout} parameter can be used to specify the number of iterations after which equality saturation should halt. The \texttt{timelimit} parameter can be used to limit the execution time by specifying a time period with the types provided by the Julia standard library module \texttt{Dates}, for example saturation time can be limited to \texttt{Seconds(30)} or \texttt{Minute(2)}. \texttt{eclasslimit} and \texttt{enodelimit} can be used to halt saturation if the egraph exceeds a certain size in terms of e-nodes and e-classes. The \texttt{goal} and \texttt{stopwhen} parameters are used to provide a custom stopping condition for the e-graph. The \texttt{goal} parameter accepts an instance of a \texttt{SaturationGoal} type, that can be constructed arbitrarily to hold a context or additional parameters. The \texttt{stopwhen} function should instead be a function object with no arguments, called after each saturation iteration, that returns true when equality saturation should stop eagerly and false otherwise. The \texttt{threaded} parameter can be used to enable multi-threading for e-graph pattern matching, while \texttt{simterm} is used to override the \texttt{similarterm} function from TermInterface (\autoref{sec:terminterface}) in the case where expression types inside the e-graph should be constructed differently from the terms used in a classical rewriting context.

\begin{figure}
	\centering
    \includegraphics[width=\textwidth]{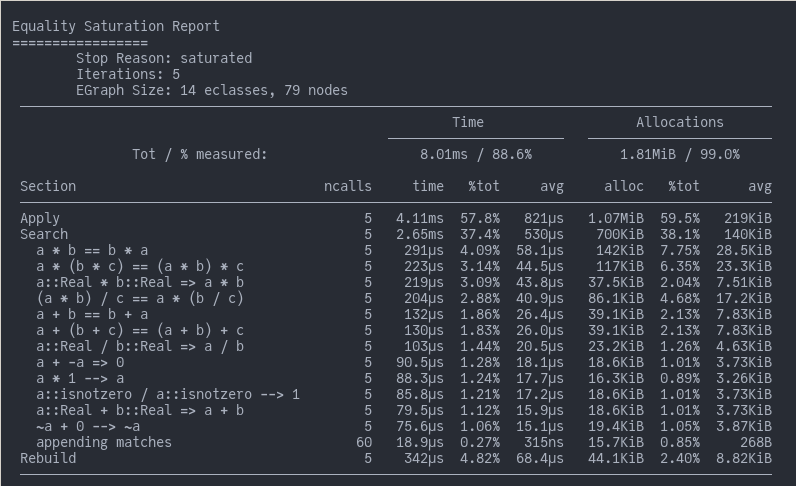}
    \caption{Report printed when executing the example code in \autoref{fig:mtexample}}
    \label{fig:eqsatreport}
\end{figure}

Our implementation of equality saturation can print a detailed execution report after execution. The execution report shows the reason why equality saturation has stopped, the size of the e-graph in terms of e-nodes and e-classes and can additionaly show a table displaying the time and memory required to search and apply each rule.
The \texttt{timer} parameter can be set to \texttt{false} to disable the rule-level measurement of timing and allocations. In \autoref{fig:eqsatreport} we show an example equality saturation report that was printed in author's Julia command line REPL interface when executing the example code in \autoref{fig:mtexample}.

\begin{figure}
    \centering

\begin{jllisting}[language=julia, style=jlcodestyle]
@with_kw mutable struct SaturationParams
    timeout::Int = 8
    timelimit::Period = Second(-1)
    matchlimit::Int = 5000
    eclasslimit::Int = 5000
    enodelimit::Int = 15000
    goal::Union{Nothing, SaturationGoal} = nothing
    stopwhen::Function = ()->false
    scheduler::Type{<:AbstractScheduler} = BackoffScheduler
    schedulerparams::Tuple=()
    threaded::Bool = false
    timer::Bool = true
    printiter::Bool = false
    simterm::Function = similarterm
end
\end{jllisting}
    \caption{Equality Saturation Parameters in Metatheory.jl}
    \label{fig:saturationparams}
\end{figure}

\FloatBarrier

\subsubsection{Equality Saturation Schedulers}
\label{ssec:schedulers}
During equality saturation, searching and applying each rule in a rewrite system may be highly inefficient. This is why Metatheory.jl and egg \cite{willsey2021egg} provide a system called \textit{schedulers} for reducing the search space in terms of applicable rules. An e-graph (or equality saturation) scheduler is a mechanism that is automatically informed about the matches produced by each rule in the rewrite system that is under consideration. A scheduler can then \textit{ban} a specific rule that is slowing down the saturation process for a certain number of iterations, based on the information it received from the matches in the previous iterations. Custom schedulers can provide substantial performance improvements, since they can control when troublesome rewrite rules can be searched and applied.  In Metatheory.jl users can define a custom scheduler type by subtyping the \texttt{Schedulers.AbstractScheduler} type and defining the methods in the schedulers interface\footnote{A detailed definition of the schedulers interface is vailable in the \texttt{src/EGraphs/Schedulers.jl} source file of \href{https://github.com/JuliaSymbolics/Metatheory.jl/}{https://github.com/JuliaSymbolics/Metatheory.jl/}}.

Egg and Metatheory.jl adopt the \textit{backoff} scheduler by default. Its policy is to identify the rules that are matching in exponentially growing locations in the e-graph and temporarily ban them.
In the exponential backoff scheduler there exists a configurable initial match limit (the \texttt{matchlimit} parameter in \texttt{SaturationParams}, as seen in \autoref{fig:saturationparams}). If a rule search yields more matches than this limit, then the rule is banned by the scheduler for a certain number of iterations. Then, the limit and the number of iterations the rule will be banned next time are doubled.

\subsection{E-Graph Pattern Matching}
\label{ssec:ematching}

E-graph pattern matching (\textit{e-matching} in short) is the procedure of searching an e-graph for all the possible matches that can be produced by a pattern. E-matching is in theory NP-hard \cite{kozen1977complexity}, and the number of matches can be exponential in the size of the e-graph. Differently from classical pattern matching, a single pattern can yield many different substitutions in different locations and those substitutions are from pattern variables to e-class ids. Thus, the e-graph pattern matching procedure requires extensive backtracking. Our implementation of the \textit{e-matching} procedure is inspired from egg's implementation \cite{willsey2021egg}, which is in turn based on the Leonardo De Moura's paper "\textit{Efficient E-matching for SMT Solvers}" \cite{de2007efficient}. In this publication, the proposed method is to use a virtual machine for e-matching. Other e-matching approaches, such as the algorithm used in the Simplify theorem prover \cite{detlefs2005simplify} have shown to be inefficient when compared to this algorithm\footnote{\href{https://www.philipzucker.com/egraph-2/}{https://www.philipzucker.com/egraph-2/}}.
In this virtual machine approach patterns are compiled to sequential lists of instructions that are used as programs for the virtual machine. We use a simplified version of this virtual machine inspired from egg, where the machine memory stores e-class ids instead of pointers to e-nodes.
Pattern variables in the original pattern are then mapped to specific registries in the virtual machine memory. If a program execution on the virtual machine terminates completely without failing, a substitution can be then extracted from the virtual machine by retrieving the contents of the registries that are mapped to the pattern variables. Our e-matching virtual machine uses an implicit backtracking stack, relying on recursive function calls and thus on Julia's internal function call stack. Our contributions include eager matching of ground terms in patterns to avoid their repeated e-matching and additional instructions to check user specified predicates and type assertions. Those type assertions can lift a literal value out of an e-class into the substitutions yielded by the pattern matcher in order to be used consistenly as a regular Julia value in the right-hand side of any \texttt{DynamicRule}.

When using the virtual machine e-matching algorithm, it is possible to highly parallelize the search phase of equality saturation. There can be many separate virtual machine instances, each residing in its own thread and each analyzing a batch of e-class ids from an e-graph. In our implementation of equality saturation, users can choose when to enable multi-threading or with a specific parameter (\autoref{fig:saturationparams}).

\subsection{E-Graph Analyses and Extraction}
\label{ssec:analyses}
With Metatheory.jl, modeling analyses and conditional/dynamic rewrites is straightforward. It is possible to check conditions on runtime values or to read and write from external data structures during rewriting. The analysis mechanism described in egg \cite{willsey2021egg} and re-implemented in this package allows users to define ways to compute additional analysis metadata from an arbitrary join-semilattice domain, such as costs of nodes or logical statements attached to terms. Other than for inspection, analysis data can be used to modify expressions in the e-graph both during rewriting steps and after e-graph saturation.
The extraction of a single term from the multitude of equivalent expressions in an e-graph can be performed by computing which e-node is the 'best-choice' for each e-class as an on-the-fly e-graph analysis. Users can define their own cost function, or choose between a variety of predefined cost functions for automatically extracting the best-fitting expressions from the equivalence classes represented in an e-graph.

\begin{defn}
\textbf{E-Graph Analysis:}\\
An \textit{e-graph analysis} (also known as e-class analysis in the original definition in \cite{willsey2021egg}) defines a join semi-lattice domain $D$ and associates a value $d_c \in D$ to each e-class $c$ in a given e-graph. Given an e-graph analysis on a domain $D$, we report the following properties of e-graph analyses from \cite{willsey2021egg}.

\begin{enumerate}
    \item It is easy to retrieve the analysis data $d_c$ associated to an e-class $c$ through a function
    $getdata : EClass \to D$, but it is otherwise hard to compute the inverse of this function.
   
    \item A function $make : ENode \to D$ is defined, computing a new value $d_c \in D$ associated to the singleton e-class $c$, which is respectively created when a new, single e-node $n$ is added to the e-graph. This function typically computes the result by accessing the analysis values $d_{c_i}$ of $n$'s children e-classes $c_i$ for $i = 1, \hdots, ar(n)$.
    
    \item A function $join : (D, D) \to D$ is defined, respectively computing the join operation of the join-semilattice domain $D$. This function should return the analysis value $d_c$ of an e-class $c$ from $d_{c_1}, d_{c_2}$ when two eclasses $c_1, c_2$ are being merged into $c$. 
   
    \item A function $modify : EClass \to EClass$ can be \textit{optionally} defined,  modifying the contents of an e-class $c$ based on $d_c$, typically by adding an e-node to $c$. $modify$ should be idempotent if no other changes occur to the e-class. For example: $modify(modify(c)) = modify(c)$.
    
    \item The previously defined operations respect a property called the \textit{analysis invariant}. Given an e-graph $G$, the data associated with each e-class in $G$ must be the join of the make for every e-node in that e-class. The second part of the invariant ensures that the modifications induced by $modify$ are driven to a fixed point.
    \begin{equation}
        \forall c \in G . \quad  d_c = \bigwedge_{n \in c} make(n) \text{ and } modify(c) = c
    \end{equation}
\end{enumerate}

\end{defn}

\subsubsection{Implementation of E-Graph Analyses}

Thanks to Julia's mechanism of multiple dispatch, defining an e-graph analysis for Metatheory.jl is a straightforward process. Differently from the original implementation in egg (\cite{willsey2021egg}), users can associate many different analyses to an e-graph by defining an abstract type that characterizes a particular analysis domain. Programmers that want to define a custom e-graph analysis are then required to dispatch the methods \texttt{make, join} and optionally \texttt{modify!} on this newly defined analysis abstract type. The internal e-graph maintenance algorithms in Metatheory.jl will take care of maintaning the invariant and associating each e-class in the e-graph to its value from the specified domain. Another feature that is available in Metatheory.jl but not in egg, is marking a particular e-graph analysis as \textit{lazy}. A \textit{lazy analysis} implies that each analysis value for each e-class is only computed \textit{on-demand}, meaning that the values will not be computed on-the-fly. If the methods \texttt{make, join} and \texttt{modify!} are computationally intensive for a given analysis domain, it is best to set the analysis as \textit{lazy}. This will allow computing \texttt{make, join} and \texttt{modify!} recursively for a specific e-class only when the users calls the \texttt{analyze!} function on the e-graph and a specific e-class id.

\section{Practical Equality Saturation in Julia}
\label{sec:usage}

In the last section of this chapter we will give some practical examples of how to use the equality saturation features in Metatheory.jl. We will briefly go through how to construct a customized e-graph analysis and how to define a custom cost function for extraction. In the last part of this section we will see how we can efficiently use equational rewrite rules to rewrite some example mathematical expressions, relying on the cost function and the analysis we constructed before. 

\subsection{Defining a Custom Analysis}
\label{ssec:customanalysis}

Here we will describe how to define custom e-graph analyses in Metatheory.jl. In this particular example we will define an analysis that will label each e-class in an e-graph with values from a custom domain that can be used to represent the sign of mathematical expressions. This analysis will become useful in \autoref{ssec:mtexample}, where we will use this information to define conditional rules with \textit{predicates}. Instead of following the precise definitions from calculus, we will define this analysis with respect to Julia's numerical operations. Evaluating \texttt{0/0} in Julia yields \texttt{NaN}, while evaluating a positive number divided by zero yields \texttt{Inf}, while for a negative numerator division by zero yields  \texttt{-Inf}.

We will tag e-classes with values

\begin{itemize}
\item \texttt{nothing} if the sign is unknown
\item 1 if the expressions represented by the e-class represent a positive value
\item -1 if they represent a negative value.
\item 0 if they represent a zero value.
\item \texttt{NaN} if they represent a \texttt{NaN} value. For example, resulting from a \texttt{0/0} operation.
\item \texttt{Inf} and \texttt{-Inf} if they represent a positive or negative infinite value.
\end{itemize}

Let's suppose that the language of the symbolic expressions that we are considering will contain only integer and real numbers, variable symbols and the \texttt{+}, \texttt{*} and \texttt{/} operations.
Since we are in a symbolic computation context, we are not interested in the actual numeric result of the expressions represented in the e-graph. Thus, we only care to analyze and identify the sign of symbolic expressions.
We will see that defining an e-graph analysis in Metatheory.jl is similar to the process of defining a procedure that involves mathematical induction. To define a custom e-graph analysis, one should start by defining a type that subtypes \texttt{AbstractAnalysis} that will be used to identify this specific analysis domain and to dispatch the analysis functions against the correct methods.

We first begin by loading the Metatheory.jl package and defining the abstract type characterizing our analysis domain.

\begin{figure}[hbt!]
    \centering
\begin{jllisting}[language=julia, style=jlcodestyle]
using Metatheory
abstract type SignAnalysis <: AbstractAnalysis end
\end{jllisting}
\end{figure}

The next step is to define a method for \texttt{make}, dispatching against our \\ \texttt{SignAnalysis} abstract type. First, we want to
associate an analysis value only to the real numbers constants contained in the e-graph. To do this we
take advantage of multiple dispatch against the type \texttt{ENodeLiteral\{T\}} concrete type, that helps the system to distinguish an e-node representing a single literal value of type \texttt{T} from an e-node representing a composite term, in turn represented by objects of type \texttt{ENodeTerm}.

\begin{figure}[hbpt!]
    \centering
\begin{jllisting}[language=julia, style=jlcodestyle]
function EGraphs.make(an::Type{SignAnalysis}, g::EGraph, n::ENodeLiteral{<:Real})
  if n.value == Inf 
      return Inf
  elseif n.value == -Inf 
    return -Inf
  elseif n.value isa Real # in Julia NaN is a Real
    return sign(n.value)
  else 
    return nothing
  end
end
\end{jllisting}
\end{figure}

\FloatBarrier

We can now define the \texttt{make} method that will dispatch against e-nodes representing composite terms. Knowing that our language contains only simple mathematical operations, with respect to the algebraic properties of these operations, we know from basic mathematics and the Julia core numerical operations that:

\begin{itemize}
    \item If $x = \texttt{NaN}$ or $y = \texttt{NaN}$, every operation $f \in \{+, \cdot, /, -\}$ yields $f(x,y) = \texttt{NaN}$, thus $-\texttt{NaN} = \texttt{NaN}$
    \item If $x$ is a real number, not $\pm \infty$ or \texttt{NaN}, then $x \cdot 0 = 0$, thus the sign is 0
    \item In the Julia arithmetics, $ \texttt{NaN} = \pm\infty \cdot 0  = \pm\infty/\pm\infty = 0/0 = +\infty - \infty = -\infty + \infty$  
    \item If $x$ is not $0$, then multiplying by -1 yields a value where the sign is also multiplied by -1.
    \item $0 \pm 0 = 0 = x/\pm\infty$ where $x$ is any non-infinite non-NaN value.
    \item The other rules from the algebra of infinites apply.
\end{itemize}

We can now define the method for \texttt{make} dispatching against 
\texttt{SignAnalysis} and \texttt{ENodeTerm} to compute the analysis value for composite symbolic terms.  We take advantage of the methods in TermInterface.jl (\autoref{sec:terminterface}) to inspect the content of an \texttt{ENodeTerm}.
From the definition of an e-node (\autoref{defn:egraphintuitive}), we also know that children of e-nodes are always e-class ids.

\begin{figure}[hbpt!]
    \centering
\begin{jllisting}[language=julia, style=jlcodestyle]
function EGraphs.make(an::Type{SignAnalysis}, g::EGraph, n::ENodeTerm)
  # Let's consider only binary function call terms.
  if exprhead(n) == :call && arity(n) == 2
      # get the symbol name of the operation
      op = operation(n)
      op = op isa Function ? nameof(op) : op 

      # Get the left and right child eclasses
      child_eclasses = arguments(n)
      l = g[child_eclasses[1]]
      r = g[child_eclasses[2]]

      # Get the corresponding SignAnalysis value of the children
      # defaulting to nothing 
      lsign = getdata(l, an, nothing)
      rsign = getdata(r, an, nothing)

      (lsign == nothing || rsign == nothing ) && return nothing

      if op == :*
        return lsign * rsign
      elseif op == :/
        return lsign / rsign
      elseif op == :+
        s = lsign + rsign
        iszero(s) && return nothing
        (isinf(s) || isnan(s)) && return s
        return sign(s)
      elseif op == :-
        s = lsign - rsign
        iszero(s) && return nothing
        (isinf(s) || isnan(s)) && return s
        return sign(s)
      end
  end
  return nothing
end
\end{jllisting}
\end{figure}

\FloatBarrier

We have now defined a way of tagging each e-node in the e-graph with its sign, reasoning inductively on the analyses values of children e-classes. The
\texttt{analyze!} function in Metatheory.jl will do the job of walking the e-graph recursively. The missing piece, is now informing Metatheory.jl about how to merge together
analysis values. Since e-classes represent many equal e-nodes, we have to inform the automated analysis algorithm how to extract a single value out of the many analyses that are constructed by \texttt{make} from the e-nodes in an e-class. We do this by defining a method for \texttt{join} for our analysis domain.

\begin{figure}[hbt!]
    \centering
\begin{jllisting}[language=julia, style=jlcodestyle]
function EGraphs.join(an::Type{SignAnalysis}, a, b)
    return a == b ? a : nothing
end
\end{jllisting}
\end{figure}

We do not care to modify the content of e-classes in consequence of our analysis.
Therefore, we can skip the definition of \texttt{modify!}.
Let's suppose that symbol \texttt{:x} will represent a positive number value, 
symbol \texttt{:y} will represent a negative value, symbol \texttt{:z} a zero value and symbol \texttt{:k} a $+\infty$ value.
Unfortunately the Julia language has no built-in mechanism of attaching arbitrary metadata to objects
of type \texttt{Symbol}, therefore this example is only valid in the context of e-graph rewriting, and we must inform the analysis mechanism of the sign of symbols by overriding \texttt{make}. We will see in \autoref{sec:symbolics} how the Symbolics.jl CAS provides extended symbol types that will allow for flexible metadata storage.

\begin{figure}[hbt!]
    \centering
\begin{jllisting}[language=julia, style=jlcodestyle]
function EGraphs.make(an::Type{SignAnalysis}, g::EGraph, n::ENodeLiteral{Symbol})
  s = n.value 
  s == :x && return 1
  s == :y && return -1 
  s == :z && return 0
  s == :k && return Inf 
  return nothing
end
\end{jllisting}
\end{figure}

We are now ready to test our analysis. 
\begin{figure}[hbt!]
    \centering
\begin{jllisting}[language=julia, style=jlcodestyle]
function custom_analysis(expr)
    g = EGraph(expr)
    analyze!(g, SignAnalysis)
    return getdata(g[g.root], SignAnalysis)
end

custom_analysis(:(3*x))        # positive, sign is 1
custom_analysis(:(3*(2+a)*2))  # nothing, we don't know the sign of a
custom_analysis(:(-3y * (2x*y))) # sign is -1
custom_analysis(:(k/k))          # is NaN
\end{jllisting}
\end{figure}

\subsubsection{Predicates and Conditional Rules}
Analyses are useful to define \textit{predicates}. Predicates help write conditional rules, and users are encouraged to use e-graph analyses to write predicates for e-graph rewriting. For example, with our newly defined \texttt{SignAnalysis} we can define the \texttt{isnotzero} predicate that returns \texttt{true} if and only if a symbolic expression is different from 0. We take advantage of multiple dispatch: if the predicate accepts two arguments, an e-graph and an e-class, then we're sure we are in an e-graph rewriting context. If we are in a classical rewriting context, then the predicate will receive only one value: the expression that has to be checked.

\begin{figure}[hbt!]
    \centering
\begin{jllisting}[language=julia, style=jlcodestyle]
# isnotzero(x::Symbol) is not possible, since we cannot attach
# metadata to Julia's built-in Symbol! We'll later use Symbolics.jl 
# for supporting predicates and metadata for classical rewriting.
# Thus, we cannot define isnotzero for composite expressions.
# We can only define
isnotzero(x::Real) = !iszero(x)
# we are cautious, so we should return false by default 
isnotzero(g::EGraph, x::EClass) = getdata(x, SignAnalysis, 0) != 0

cansimplifyfraction(x::Real) = !iszero(x) && !isnan(x) && !isinf(x)
function cansimplifyfraction(g::EGraph, x::EClass)
    sign = getdata(x, SignAnalysis, 0) 
    sign == nothing && return false
    return cansimplifyfraction(x)
end

# A conditional rewrite
@rule ~a::cansimplifyfraction / ~a::cansimplifyfraction --> 1 
\end{jllisting}
\end{figure}

\FloatBarrier

\subsection{Defining a Custom Cost Function for Extraction}
\label{ssec:costfun}

Extraction of a single term from an e-graph, representing possibly infinitely many equivalent symbolic expressions, is formulated as an \textit{e-graph} analysis.
A cost function for e-graph extraction is a function used to determine which e-node will be extracted from an e-class containing multiple equivalent nodes. 
The default cost function used for extraction is \texttt{astsize}, which measures the size of expressions recursively and causes the extraction analysis to adopt the tendency of choosing smaller terms over bigger terms. An e-graph cost function must return a positive, non-complex number value and must accept 3 arguments:

\begin{enumerate}
    \item The \texttt{ENode} \texttt{n} that is being inspected.
    \item The \texttt{EGraph} \texttt{g} that is being analyzed.
    \item The current analysis type \texttt{an}. The system will take care of automatically creating an abstract type to represent the extraction analysis, and this argument can be used to access the data of \texttt{n}'s children.
\end{enumerate}

From those 3 parameters, one can access all the data needed to compute the cost of an e-node recursively. One can use TermInterface.jl methods to access the operation and child arguments of an e-node: \texttt{operation(n)}, \texttt{arity(n)} and \texttt{arguments(n)}.
Since children of e-nodes always point to e-classes in the same e-graph, one can retrieve the list of  \texttt{EClass} objects for each child of the currently visited e-node with \texttt{[g[id] for id in arguments(n)]}.
One can then inspect the analysis data for a given e-class and a given extraction analysis type \texttt{an}, by using the functions \texttt{hasdata} and \texttt{getdata}.
Extraction analyses always associate a tuple of 2 values to a single e-class: the first element is the e-node in that class that minimizes the cost, the second element of the tuple is the actual cost of that node.
In \autoref{fig:costfun} we show the code for an example cost function that behaves like \texttt{astsize} but increments the cost of nodes containing the \texttt{*} operation. This results in a tendency to avoid 
extraction of expressions containing \texttt{*}. In this example cost function, we want all literal values in the e-graph to have cost 1, but we want to compute the cost of composite terms inductively. The extraction mechanism is easily extensible by defining a single function, using multiple dispatch and simple TermInterface.jl methods, without forcing the library users to define a completely new analysis for the e-graph. 

\begin{figure}[hbt!]
    \centering
\begin{jllisting}[language=julia, style=jlcodestyle]
function cost_function(n::ENodeTerm, g::EGraph, cost_analysis)
    cost = 1 + arity(n)

    nameof(operation(n)) == :* && (cost += 2)

    for id in arguments(n)
        eclass = g[id]
        # if the child e-class has not yet been analyzed, return +Inf
        !hasdata(eclass, cost_analysis) && (cost += Inf; break)
        cost += last(getdata(eclass, cost_analysis))
    end
    return cost
end

cost_function(n::ENodeLiteral, g::EGraph, cost_analysis) = 1
\end{jllisting}
\caption{Definition of a custom cost function for e-graph extraction}
\label{fig:costfun}
\end{figure}

\FloatBarrier

\subsection{Complete Example of E-Graph Rewriting}
\label{ssec:mtexample}

In this complete example (\autoref{fig:mtexample}), we build a collection of rewrite systems, called \textit{theories} in Metatheory.jl, with the aim of simplifying expressions in the usual commutative monoid of multiplication, the commutative group of addition and some properties of division. We compose the \textit{theories} together
with a \textit{constant folding} theory. The pattern matcher for the e-graphs backend
allows us to use the existing Julia type hierarchy for integers and
floating-point numbers with a high level of abstraction to define rules that compute constant folding. As a contribution over
the original egg \cite{willsey2021egg} implementation, left-hand sides of rules
in Metatheory.jl can contain predicate checks and type assertions on pattern variables, allowing programmers to define conditional rules that depend on consistent type hierarchies and to seamlessly access literal
Julia values in the right-hand side of dynamic rules.
We finally introduce two simple rules for simplifying fractions that check the additional analysis data we defined in \autoref{ssec:customanalysis}.
\autoref{fig:eqsatpic} contains a friendly visualization of a consistent fragment of the equality saturation process in this example.
It is easy to see how loops in computations would evidently arise from classical rewriting misuse of the set of rewrite rules that we provide in this example.
While the classic rewriting backend would loop indefinitely or stop early when repeatedly matching these rules, and thus requires the user to either change the system or provide a strategy using functional combinators (\autoref{ssec:combinators}),
the e-graph backend natively supports this level of abstraction and allows the
programmer to define equational rules and completely forget about the ordering and looping of rules.
Efficient scheduling heuristics are applied automatically to prevent instantaneous
combinatorial explosion of the e-graph, thus preventing substantial slowdown of the equality saturation
process.

\begin{figure}[H]
    \centering
\begin{jllisting}[language=julia, style=jlcodestyle]
using Metatheory
using Metatheory.EGraphs

# pattern variables can be specified before the block of rules
comm_monoid = @theory a b c begin  
  a * b == b * a # commutativity
  a * 1 --> a    # identity
  a * (b * c) == (a * b) * c   # associativity
end;

# theories are just vectors of rules
comm_group = [
  @rule a b (a + b == b + a) # commutativity
  # pattern variables can also be written with the prefix ~ notation
  @rule ~a + 0 --> ~a   # identity
  @rule a b c (a + (b + c) == (a + b) + c)   # associativity
  @rule a (a + (-a) => 0) # inverse
];

# dynamic rules are defined with the `=>` operator
folder = @theory a b begin
  a::Real + b::Real => a+b
  a::Real * b::Real => a*b
end;

div_sim = @theory a b c begin
  (a * b) / c == a * (b / c)
  a::cansimplifyfraction / a::cansimplifyfraction --> 1 
end;

t = vcat(comm_monoid, comm_group, folder, div_sim) ;

g = EGraph(:(a * (2*3) / 6)) ;
analyze!(g, SignAnalysis)
saturate!(g, t) ;
ex = extract!(g, astsize) # will result in :a
\end{jllisting}
    \caption{Example usage of Metatheory.jl}
    \label{fig:mtexample}
\end{figure}

\chapter{Applications and Results}
\label{ch:applications}

In this chapter we will discuss how the powerful Julia homoiconicity and meta-language features will allow programmers to use our optimization framework to rewrite any Julia program, achieving performance improvements by transforming programs into equivalent optimized forms before evaluating the code. With this approach, programmers can provide a high-level equational theory directly in the same programs that have to be optimized. We will then review some results about how the symbolic computation packages and the expression rewriting framework introduced in \autoref{ch:solution} can benefit Julia programmers and scientists. We will focus mostly on the integration of the rewriting features of Metatheory.jl into the Julia CAS Symbolics.jl \cite{gowda2021high}, and how this integration can help scientists to write domain-specific compiler optimizations for large symbolic-numeric simulations with ModelingToolkit.jl \cite{ma2021modelingtoolkit}.

\section{Example of Functional Stream Fusion}
\label{sec:streamfusion}

\subsection{Loop Fusion}

Stream fusion is the practice of optimizing iterator expressions in functional programming languages \cite{coutts2007stream, hinze2010theory}.
The Julia language already provides built-in conventient syntactical loop fusion.
The \texttt{@.} macro can take an expression and convert every function call and operator in the argument expression returning a new expression with vectorized (also known as \textit{broadcasted}) code. Broadcasting is the operation of applying a function $f$ over the elements of sized containers and scalars and returning an appropriately sized container (or scalar) as the result. Broadcasting is recognized by Julia on a syntactical level: a broadcasted function call \texttt{f.(X)} can be distinguished from a regular function call \texttt{f(X)} by the use of the prefix dot. Conveniently, dots are supported for binary operators, for example \texttt{A .+ B} returns the broadcasting of \texttt{+} over the two containers \texttt{A} and \texttt{B}. \texttt{A} could be a $5 \times 1$ vector and  \texttt{B} a $5 \times 2$ matrix, and the result would be a $5 \times 2$ matrix obtained by summing every column of \texttt{B} with the vector \texttt{A}. Broadcasted operations can be seen as a syntactical feature: the dotted operators and call syntax are syntactical sugar for \texttt{broadcast(f, arguments...)}. Thus, the dots allow Julia to recognize the "vectorized" nature of the operations at a syntactic level and to perform loop fusion as a language feature instead than as a compiler optimization. For example, the expression \texttt{3 .* x .+ y} is equivalent to \texttt{(+).((*).(3, x), y)} and it can be syntactically fused into the expression \texttt{broadcast((x,y) -> 3*x+y, x, y)} before even lowering code to intermediate IRs and applying compiler optimizations.
This syntax-level loop fusion indeed a very smart language feature that can help programmers write very fast vectorized code in a severely concise way.

The Julia language though, also provides functional programming streams such as the \texttt{map, filter} and \texttt{reduce} higher-order functions, common to almost all functional programming languages. Those functions accept callable objects and one or more iterable collections and apply the argument callable objects in different ways on the elements of iterable collections. Are those functional stream loops fused syntactically by Julia? The answer is no, since there is no built-in language construct such as dotted operators, and there is \textit{no} guarantee that the functions passed as arguments to manipulate the collections are functionally pure. As a practical, small-sized example of domain-specific compiler optimizations we are showing the implementation of a toy functional stream optimizer on the syntax level, with the strong assumption that we are going to use this high-level optimizer only on purely functional code.

Although the behaviour of functional streams can often concide with broadcasted code (see for example \texttt{sqrt.([1,2,3]) == map(sqrt, [1,2,3])}), there are substantial differences between \textit{broadcasting} and functional streams:
broadcasting is designed to handle shaped containers such as tensors, vectors and matrices, while functional streams such as \texttt{map} handle shapeless collections like iterators and sets. Functional streams tend to treat their arguments as containers by default, while broadcasting treats its arguments as scalars by default. One has to define a container to be explicitly broadcastable, and most importantly functional streams such as \texttt{map}
cannot combine arrays and scalars, whereas broadcasting can expand smaller containers to match larger ones
\footnote{See this blog post for more details about broadcasting and the difference with functional streams: \href{https://julialang.org/blog/2017/01/moredots/}{https://julialang.org/blog/2017/01/moredots/}}. 
Thus, broadcasting is designed for sized containers such as matrices and vectors and allows SIMD optimization, while functional streams in Julia are more appropriate to manipulate streams of data of unknown size or \textit{shapeless} iterators.

In this example application \footnote{Thanks to Philip Zucker (\href{https://www.philipzucker.com/}{https://www.philipzucker.com/}) for the idea and the initial implementation} we are going to show the power of Metatheory.jl (\autoref{sec:metatheory}) by implementing a simple expression-level optimizer that fuses functional streams in the assumption of functional purity. 
Note that functional purity of Julia code is an extreme assumption, and this optimizer is thus incomplete and ratherly naive, and must only be seen as an example application for optimizing domain-specific code. Since Julia supports functional programming primitives but is not purely functional, in order to achieve the correctness of the results this stream fusion optimizer should check the purity of the functions that are used to map and filter. By using impure functions that mutate state, the results of this optimizer will much likely be code with undefined behaviour. When loop fusion is viewed as a compiler optimization, the compiler should be able to fuse only if it can prove that the optimization won't alter results, which requires extensive static analysis of the code for detecting impure code, a lower-level compiler topic that we will not cover here. The goal of this example is to show how practical complex compiler optimizations can be defined in a high-level syntax that resembles equational algebraic theories in mathematics. Thanks to Metatheory.jl and Julia's powerful metaprogramming features, this kind of optimization can reside directly inside of the program that has to be optimized.

\subsection{Implementation of the Functional Stream Optimizer}

We begin defining our custom optimizer by giving the equational axioms in \autoref{fig:streamrules} that dictate the relations between functional stream expressions.
In order to support this kind of equational reasoning in a functional programming context, we can use two new function symbols that will \textit{not be defined} in the program or Julia's standard library, but will only be used as a syntactical construct to be rewritten later in another step of the optimizer. Those expressions are \texttt{apply(f, x)} that will be rewritten appropriately to \texttt{f(x)} and the expression \texttt{fand(f,g)} that represents the functional \textit{AND} operation for single-argument functions, and will be rewritten to \texttt{(x -> f(x) \&\& g(x))}. 

\begin{figure}[ht]
    \centering
\begin{jllisting}[language=julia, style=jlcodestyle]
stream_theory = @theory x y f g M N begin
    map(f,fill(x,N))            == fill(apply(f,x), N)
    fill(x,N)[y]                --> x
    length(fill(x,N))           --> N
    reverse(reverse(x))         --> x
    sum(fill(x,N))              --> x * N
    map(f,reverse(x))           == reverse(map(f, x))
    filter(f,reverse(x))        == reverse(filter(f,x))
    reverse(fill(x,N))          == fill(x,N) 
    filter(f, fill(x,N))        == (apply(f, x) ? fill(x,N) : fill(x,0))
    filter(f, filter(g, x))     == filter(fand(f,g), x)
    cat(fill(x,N),fill(x,M))    == fill(x,N + M)
    cat(map(f,x), map(f,y))     == map(f, cat(x,y))
    map(f, cat(x,y))            == cat(map(f,x), map(f,y)) 
    map(f,map(g,x))             == map(f ∘ g, x)
    reverse( cat(x,y) )         == cat(reverse(y), reverse(x))
    map(f,x)[y]                 == apply(f,x[y])
    apply(f ∘ g, x)             == apply(f, apply(g, x))
    reduce(g, map(f, x))        == mapreduce(f, g, x)
    foldl(g, map(f, x))         == mapfoldl(f, g, x)
    foldr(g, map(f, x))         == mapfoldr(f, g, x)
end
\end{jllisting}
    \caption{Rewrite rules for functional stream fusion.}
    \label{fig:streamrules}
\end{figure}

On line 5 of \autoref{fig:streamrules} we have a rewrite specifying that \texttt{reverse} is its own inverse. This mean that reversing an iterator twice will yield the original iterator again. Rewrite on line 7, for example, defines the commutativity of \texttt{map} and \texttt{reverse}. Some practical optimizations happen in the rewrite on line 11. Instead of filtering a container twice for two predicates $f$ and $g$, this rewrite asserts that this can be done by iterating the container once and checking for both predicates at the same time. This rewrite can substantially reduce the time required for composite filtering of substantially big containers. Another optimization happens in the rewrite on line 17. Mapping a function $f$ over an iterator $x$ and then immediately retrieving the value at index $y$ is equivalent to applying the function $f$ directly to the element of $x$ in position $y$. This optimization can save the cost of iterating through an entire collection. In our example expressions we are going to use the function \texttt{fill(x,y)}, that creates an array of length \texttt{y} and fills it with the element \texttt{x}. For the sake of simplicity we are only considering arrays of numbers, thus, rewrite on line 6 of can ensure that the sum of an array of many identical elements is just equal to the length of that array times the element that is repeated. We can then completely spare the runtime from allocating an array in this particular case. The function \texttt{cat} concatenates two flat collections. Rewrite on line 13 asserts the commutativity of concatenating collections and mapping a function $f$.
We can now give some rules in \autoref{fig:foldtheory} for basic number constant folding that we'll use later when completing our optimizer. 

\begin{figure}[ht]
    \centering
\begin{jllisting}[language=julia, style=jlcodestyle]
fold_theory = @theory x y z begin 
    x::Number * y::Number => x*y
    x::Number + y::Number => x+y
    x::Number / y::Number => x/y
    x::Number - y::Number => x/y
end
\end{jllisting}
    \caption{Rewrite rules for simple number constant folding.}
    \label{fig:foldtheory}
\end{figure}

The Julia standard library provides a function that inlines single-argument anonymous function application expressions, providing very simple partial evaluation capabilities through substitution. Given an homoiconic expression \texttt{funex} of type \texttt{Expr} that represents an anonymous lambda function, accepting a single formal parameter \texttt{x}, and given an actual parameter \texttt{y}, the \texttt{inlineanonymous(funex, y)} function returns the body of the lambda function expression \texttt{funex} with \texttt{x} substituted to \texttt{y}.
In \autoref{fig:inlinewrap} we define a safe wrapper of this simple inlining partial-evaluation to be used as a functional rewriting combinator (\autoref{ssec:combinators}), effectively giving our simple optimizer the power of an explicit substitution calculus \cite{curry1958combinatory}. By definition, a Metatheory.jl rewriter combinator should return \texttt{nothing} if no change occured to the input expression, and should return the modified expression otherwise. We can thus define our wrapper to apply the anonymous function inlining only to anonymous function application expressions or return \texttt{nothing} otherwise.

\begin{figure}[ht]
    \centering
\begin{jllisting}[language=julia, style=jlcodestyle]
tryinlineanonymous(x) = nothing
function tryinlineanonymous(ex::Expr)
    exprhead(ex) != :call && return nothing
    f = operation(ex)
    # only accept lambdas
    (!(f isa Expr) || exprhead(f) !== :->) && return nothing
    try 
        return inlineanonymous(f, arguments(ex)[1])
    catch e 
        return nothing 
    end
end
\end{jllisting}
    \caption{Rewrite combinator wrapper for anonymous function inlining.}
    \label{fig:inlinewrap}
\end{figure}

\FloatBarrier

The last rewrites needed in order to complete our simple functional stream fusion optimizer are rules that convert expressions of the form \texttt{apply(f, x)} and \texttt{fand(f, g)} back into a form that is accepted by the Julia compiler, without having to define actual functions that compute these expressions at runtime. They are depicted in \autoref{fig:normalizetheory}.

\begin{figure}[ht]
    \centering
\begin{jllisting}[language=julia, style=jlcodestyle]
normalize_theory = @theory x y z f g begin
    fand(f, g)  => Expr(:->, :x, :(($f)(x) && ($g)(x)))
    apply(f, x) => Expr(:call, f, x)
end
\end{jllisting}
    \caption{Rewrite rules for converting functional helper expressions back into the Julia form.}
    \label{fig:normalizetheory}
\end{figure}

We can now combine those building blocks together into a single function that given a Julia expression \texttt{ex} returns the optimized form. We show the code in \autoref{fig:streamoptimizer}. With Metatheory.jl we can elegantly compose the power of e-graph rewriting together with classical rewriting in a very high-level fashion. Our optimizer first applies the equational rewrites that define the relations between functional streams in Julia through equality saturation, extracts the shortest equivalent expression out of the e-graph and then transforms the resulting expression with classical rewriting techniques. The functional combinator used to define the strategy for classical rewriting (\autoref{ssec:combinators}), takes a \texttt{Chain} of our rules to inline anonymous functions, fold constants and convert special expressions back into the Julia form, it then walks the AST of the expression through \texttt{Postwalk} and tries to repeat these steps with \texttt{Fixpoint} until a fixed point is reached and the expression can no longer be transformed.
Since our optimizer rewrites homoiconic Julia expressions, it is indeed possible to optimize expressions in a program at compile time through the flexible Julia \textit{macro system}.
In the last lines of \autoref{fig:streamoptimizer} we define the \texttt{@stream\_optimize} macro, accepting a piece of Julia code and returning the optimized version. Differently from functions, Julia macros are executed at compile time and can be used to manipulate parts of a program, interpolating the rewritten expressions back into the program to be executed at runtime \footnote{More details about the macro system can be found at \href{https://docs.julialang.org/en/v1/manual/metaprogramming/}{https://docs.julialang.org/en/v1/manual/metaprogramming/}}.

\begin{figure}[ht]
    \centering
\begin{jllisting}[language=julia, style=jlcodestyle]
classical_rewrites = [
    tryinlineanonymous, 
    normalize_theory..., 
    fold_theory...
]

function stream_optimize(ex)
    g = EGraph(ex)
    report = saturate!(g, stream_theory, params)
    ex = extract!(g, astsize)
    ex = Fixpoint(Postwalk(Chain(classical_rewrites)))(ex)
    return ex
end

macro stream_optimize(ex)
    stream_optimize(ex)
end
\end{jllisting}
    \caption{Definition of our complete functional stream optimizer.}
    \label{fig:streamoptimizer}
\end{figure}

Note that extracting expressions by minimizing the size of the AST will not always yield an expression with an improved overall time complexity. Extracting by AST size can give satisfying results in an example setting, but is not appropriate for actually optimizing functional stream expressions. A future improvement to this example optimizer could be defining an e-graph analysis that allows extraction by minimizing an estimate of the asymptotic complexity of stream expressions rather than extraction by minimizing the size of the AST. We leave this task as an exercise for interested readers.
We can now test our optimizer by providing some example homoiconic Julia expressions. In \autoref{fig:streamexamples} we show how it can optimize some trivial functional stream expressions. The first expression
that we optimize is creating an array of length 4 filled with the number 3, then mapping a function that multiplies its argument by 7 over this array. The call to \texttt{map} can be avoided and we can thus optimize by inlining the application of the anonymous function by using rewrite on line 2 of \autoref{fig:streamrules}. Thus, the optimized expression just constructs an array of length 4 filled with the value $3 \cdot 7 = 21$ obtained by lambda inlining and partial evaluation.
The second expression that we can test is the same as the first, with the difference that the last operation performed is accessing the resulting array at position 1. We can thus completely avoid the array computation and just return 21. We then show how the \texttt{@stream\_optimize} macro we defined can be used to do this optimization at compile time to optimize the same program where the high-level domain specific optimizations were defined.

\begin{figure}[ht]
    \centering
\begin{jllisting}[language=julia, style=jlcodestyle]
# optimizing expressions at runtime
ex = :( map(x -> 7 * x, fill(3,4)))
opt = stream_optimize(ex)
@test opt == :(fill(21, 4))

ex = :( map(x -> 7 * x, fill(3,4) )[1])
opt = stream_optimize(ex)
@test opt == 21

# optimizing the expression at compile time, will return 21 at compile time
@stream_optimize map(x -> 7 * x, fill(3,4))[1] 
\end{jllisting}
    \caption{Some test cases for our functional stream optimizer.}
    \label{fig:streamexamples}
\end{figure}

\section{The Symbolics.jl Computer Algebra System}
\label{sec:symbolics}

Symbolics.jl\footnote{Source code is available at \href{https://github.com/JuliaSymbolics/Symbolics.jl}{https://github.com/JuliaSymbolics/Symbolics.jl}} \cite{gowda2021high} is a fast and modern Computer Algebra System for the Julia programming language. The goal of this CAS is to have a high-performance and parallelized symbolic algebra system that is directly extendable in the same language as the users. Since Symbolics.jl provides specialized symbolic expression types and methods for Julia generic functions in the standard library, the multiple dispatch mechanism will allow any mathematical function in Julia to work seamlessly with symbolic expressions. Users of Symbolics.jl can for example construct matrices of symbolic expressions and multiply them as they would regularly do in Julia to find out that this works out of the box. 

Symbolics.jl is built on top of SymbolicUtils.jl \footnote{Source code is available at \href{https://github.com/JuliaSymbolics/SymbolicUtils.jl}{https://github.com/JuliaSymbolics/SymbolicUtils.jl}}. The latter package provides  specialized types and systems of rewrite rules that are the core of Symbolics.jl. Symbolics.jl builds off of SymbolicUtils.jl to extend it to a whole symbolic algebra system, complete with support for differentation, solving symbolic systems of equations, fast automated sparsity detection, generation of sparse Jacobian and Hessians, and a plethora of other features.
Using Symbolics.jl, one can generate Julia code from symbolic
expressions at run time, just-in-time compile them, and then execute
them in the same session. This type of metaprogramming is much more
convenient for mathematical code than macro-based metaprogramming. Symbolics.jl
exposes a generic \texttt{toexpr} function that performs code generation by turning specialized mathematical expressions into executable homoiconic Julia code.

\subsection{Generality without sacrificing performance}
\label{ssec:symbolics_generality}

The Symbolics.jl CAS provides specialized symbolic expression types that perform fast and implicit canonicalization at the \textit{constructor level}, while still retaining generality.  The choice of term representation can surely impose restrictions on
generality. For example, the simplest term representation can be found
in Lisp-like systems or in Julia's built-in \texttt{Expr} types: terms are quoted expressions, which, in turn, are simply lists of lists or atoms. While this is an elegant and simple solution, it has a few shortcomings:
The user can, at best, define one overloading of \(+\) and \(*\) for all
expressions, but expressions can and should be of different types. The second important flaw is that the size of a term will grow with the number of operations
unless the built terms are simplified during construction, which can be
expensive using list-like data structures.

A solution to the first shortcoming is to use a parameterized structure
\texttt{Term\{T\}(f, args)} to encode symbolic terms. We can define methods for mathematical operators such as \texttt{+} and \texttt{*} that dispatch on the \texttt{Term\{<:Number\}} type and leave them open to be extended for non-number terms.
This allows users specialize the symbolic behavior in a manner that is
dependent on the type of object being acted on. 
To solve the second shortcoming, Symbolics.jl adopts a number of
constructor-level simplification mechanisms. Multiplication and addition
of numbers are the most common operations, yet simplifying commutativity
and associativity in a rule-based system can take a long time. Instead, Symbolics.jl adopts an idea from SymEngine \footnote{\href{https://symengine.org/design/design.html}{https://symengine.org/design/design.html}} that consists in formulating a canonical form in
terms of \texttt{Add} and \texttt{Mul} types, which simplify expressions upon
construction. 
\texttt{Add} represents a linear combination of terms and stores
the numeric coefficients and their corresponding terms in a key-value dictionary, respectively. \texttt{Mul} stores a product of factor
terms by storing the base terms and the corresponding exponents in a key-value dictionary. This allows us to use \(O(1)\) dictionary lookups to simplify repeated addition and multiplication. In the best case those structures can take up \(O(1)\) space, while a Lisp-like \texttt{Term} would take \(O(n)\) space for \(n\) operations.
This constructor level canonicalization provides a substantial speedup.
In a synthetic benchmark \cite{gowda2021high} which generates a random expression of 1400 terms using \(+\) and \(*\), we got a speed up of 113 \(\times\) as compared to rule-based simplification.

Having introduced these types, one may think that the generality
provided by a common term type gets lost. However, this generality is regained by adopting the set of generic functions defined by TermInterface.jl, described in \autoref{sec:terminterface}.
A new feature introduced in the Symbolics.jl system is to use more specialized types for representing polynomials in a way that is even more efficient than using the \texttt{Add} and \texttt{Mul} types described above.
The \texttt{PolyForm} type holds a polynomial in a very efficient representation while also holding the mappings necessary to present the polynomial as a valid TermInterface.jl expression. In this way, very efficient symbolic polynomial manipulation algorithms, such as fast algorithms for computing polynomial fraction simplification can be used without losing the generality introduced by TermInterface.jl.

\subsection{Metatheory.jl and Symbolics.jl}
\label{ssec:symbolics_architecture}

In \autoref{sec:architecture} we explained how we completely redesigned the architecture of the Julia symbolic computation ecosystem in order to move all the generic term-rewriting features of the packages SymbolicUtils.jl and Symbolics.jl into a package that provides a highly generic expression rewriting solution, namely Metatheory.jl, described in \autoref{sec:metatheory}. This way, any package that needs term rewriting capabilities on specialized expression types, such as Symbolics.jl, can rely on a shared symbolic expression specification provided by TermInterface.jl (\autoref{sec:terminterface}). One of the goals of this project was to highly generalize term rewriting utilities to support rewriting both mathematematical expressions from Symbolics.jl and executable homoiconic Julia code. Metatheory.jl and Symbolics.jl now share the same rule definition language, and thus the same rewrite rules and functional combinators can be used to manipulate both Symbolics.jl expressions and Julia \texttt{Expr}s. 

The main advantage of this architectural rethinking, though, is that Symbolics.jl expressions can now be manipulated by our generic optimization framework through e-graphs and equality saturation, described in \autoref{sec:metatheory} and \autoref{sec:streamfusion}.
Since Julia is a language built with technical and numerical computing in mind, Symbolics.jl expressions are often used to generate fast executable Julia code that evaluates the expressions numerically, opening a world of opportunities in symbolic-numeric computing. 
Many users of Symbolics.jl take advantage of the extensibility of the CAS through multiple dispatch to extend it with their own, domain-specific algebras and expression types.
Some examples can be found in the Julia package QuantumCumulants.jl \cite{plankensteiner2021quantumcumulants}, where Symbolics.jl is extended in order to support the symbolic derivation of mean-field equations for quantum mechanical operators, or the package Catalyst.jl \cite{ma2021modelingtoolkit, rackauckas2017differentialequations} for analysis and high performance simulation of chemical reaction networks.
Many of these packages are built to allow users to elegantly construct models of various domain-specific systems in a high-level symbolic representation and then use automatic transformations of symbolic expressions and performant code generation to later simulate models with automated, high-performance numerical solvers.

Our proposed domain-specific optimization framework can be used by programmers and scientists to pre-compute and partially evalaute Symbolics.jl expressions before feeding them into numerical simulations. Users of Symbolics.jl can provide a domain-specific system of equational axioms, a cost function and let equality saturation from Metatheory.jl (\autoref{sec:metatheory}) do the job of automatically choosing what can be the most performant equivalent symbolic expression to compile and evaluate numerically. The possibilities of this approach and the Symbolics-Metatheory integration are many:
\begin{itemize}
    \item Allow users to define symbolic rewrites, equational axioms and compiler optimizations with the same rule definition language and in the same programs.
    \item Functionally compose the systems of equational rewrite rules to achieve compositionality of the domains of reasoning, achieving fine-grained separation of concerns between symbolic implementations of different mathematical theories.
    \item Re-purpose e-graphs to prove the equality of two or more symbolic expressions or to symbolically solve equalities and inequalities in a specific domain (more details about future extensions can be found in \autoref{ch:future}).
\end{itemize}

In the next section we will briefly describe an important Julia package for symbolic-numeric simulations and we'll go through a few real world applications where we obtained substantial speedups in symbolic-numeric simulations, by using our domain-specific high-level expression optimization approach.
Our framework is particulary suitable for very large symbolic expressions and systems of differential equations that have been automatically generated by intricate models, such as simulations of robots (\autoref{ssec:robot}) or chemical reaction networks (\autoref{ssec:catalyst}). We will then see in \autoref{ch:future} how the equality saturation framework can be extended to go beyond optimization tasks, achieving domain-specific theorem proving superpowers.

\section{Symbolic-Numerics and ModelingToolkit.jl}
\label{sec:mtk}

ModelingToolkit.jl\footnote{Source code available at \href{https://github.com/SciML/ModelingToolkit.jl}{https://github.com/SciML/ModelingToolkit.jl}} \cite{ma2021modelingtoolkit} is a causal and acausal modeling framework for automatically parallelized  high-performance symbolic-numerics and scientific machine learning (SciML\footnote{See the SciML Open Source Software organization: \href{https://sciml.ai}{https://sciml.ai}}) in Julia.
ModelingToolkit.jl was created to automate the application of symbolic optimizations to large-scale numerical simulations. It was built to let users give a high-level description of a model, usually in terms of systems of differential equations, and then automatically apply symbolic optimizations before generating highly efficient numerical code, to be used in advanced numerical solvers such as DifferentialEquations.jl \cite{rackauckas2017differentialequations}. ModelingToolkit.jl can automatically generate fast executable Julia functions for symbolic model components like Jacobians and Hessians, along with automatically sparsifying and parallelizing the computations.  The conversion can also happen the other way around, allowing users to convert numerical models into symbolic models. 
Thus, ModelingToolkit.jl combines some of the features from symbolic computing with the idea of equation-based modeling systems. This package involves the ability to use the entire Symbolics.jl Computer Algebra System as part of the modeling process for integrated manipulation of symbolic expressions, in order to achieve automated transformations, simplification and compositions of models, such as index reduction, alias elimination and tearing of non-linear systems for efficient numerical handling of large-scale systems. Precisely, all the expressions defining the models and systems of differential equations in ModelingToolkit.jl are all Symbolics.jl expressions, and all the intermediate representations used by ModelingToolkit.jl are also Symbolics.jl expressions. With the new architectural rethinking of the packages in the Julia symbolic computation ecosystem, introduced in \autoref{sec:architecture}, all the e-graph rewriting and optimization features introduced in \autoref{sec:metatheory} are made directly available to be used by end users to define domain-specific optimizations for their advanced symbolic-numeric simulations, running on top of ModelingToolkit.jl

Let's leave the context of Julia for a moment, to talk about systems like Herbie \cite{panchekha2015automatically} and SPORES \cite{wang20spores}. Those compilier optimization systems use general-purpose e-graph rewriting solutions like egg \cite{willsey2021egg} to optimize programs, for example by improving numerical stability or speeding up linear algebra computations. Like in our proposed approach, these optimizations happen in a semi-automated fashion and by the same equality saturation algorithm that we use in the Metatheory.jl e-graph rewriting module. 
However, these systems currently require programmers to build sophisticated domain-specific and scalable code optimizers, and often suffer from the two-language problem. Since writing fast numerical code is not the domain of humans but is instead in the domain of symbolic engines, ModelingToolkit.jl can benefit from Symbolics.jl and the underlying Metatheory.jl rewriting system, to use classical rewriting and our proposed equational optimization framework to apply domain-specific mathematical code optimizations in an automated fashion, all in a high-performance programming language.

In a few words, our e-graph based symbolic optimization system can be used to instruct ModelingToolkit.jl to always preemptively choose the symbolic representation of a system of differential equations that performs best when compiled and solved numerically, according to an user-defined system of equational axioms. This can greatly reduce the time required to run large scale simulations through this advanced modeling framework.
This can be achieved by scientists using ModelingToolkit.jl by defining theories of equational rewrite rules that can elegantly represent the mathematical axioms of the scientific domain they are working in, all in the same program where they are coding the symbolic-numeric simulations.  In the following subsections we show how we used our e-graph rewriting framework to define domain-specific optimizations for symbolic-numeric simulations in ModelingToolkit.jl, and we show how these optimizations can substantially improve the performance of advanced numerical computations by preemptive manipulation of symbolics expressions.

\subsection{Symbolic-Numeric Simulation of Robot Dynamics}
\label{ssec:robot}

\begin{figure}[ht]
    \centering
    \includegraphics[width=.6\textwidth]{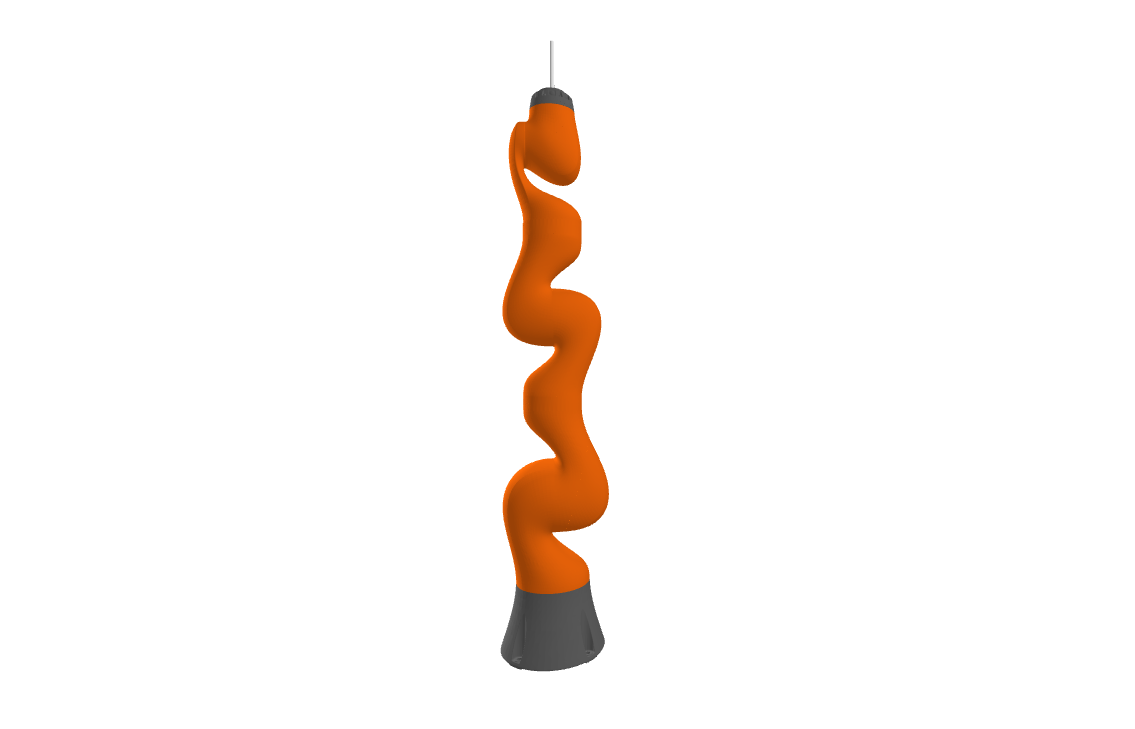}
    \caption{The KUKA IIWA 14 robotic arm 3D model visualized with MeshCat.jl}
    \label{fig:robot}
\end{figure}

Symbolics.jl provides an overall speedup in symbolic computations when compared
to other solutions, such as SymPy \cite{meurer2017sympy}. 
In our paper "High-performance symbolic-numerics via multiple dispatch" \cite{gowda2021high}
we reported how we obtained a 2370$\times$ speedup when tracing into the symbolic mass matrix computation of the rigid body dynamics system with 7 degrees of freedom used to simulate a KUKA IIWA 14 \cite{selic2013robot} robotic arm, displayed in \autoref{fig:robot}. The simulation used the packages ModelingToolkit.jl \cite{ma2021modelingtoolkit}, RigidBodyDynamics.jl \footnote{Source code available at \href{https://github.com/JuliaRobotics/RigidBodyDynamics.jl}{https://github.com/JuliaRobotics/RigidBodyDynamics.jl}} \cite{koolen2019julia}, and TORA.jl \footnote{Source code available at \href{https://github.com/JuliaRobotics/TORA.jl}{https://github.com/JuliaRobotics/TORA.jl}} \cite{torajl}. The result we published though, didn't consider the time required to evaluate the mass matrix numerically but only considered the time required for Symbolics.jl to construct the symbolic mass matrix expressions.

Here, we show how using our optimization framework to symbolically optimize the mass matrix of rigid body dynamic systems can provide substantial performance improvements in later numerical evaluations of the matrix. Our system applies a compiler optimization specifically tailored for this computation. We defined a symbolic partial evaluator that prunes unnecessary branches of the mass matrix symbolic syntax tree through equality saturation (\autoref{ssec:eqsat}). The partial evaluator first builds an e-graph out of the symbolic mass matrix and then applies the equational rewrites representing commutativity, associativity and distributivity of multiplication and addition.
AST nodes of the form $x \cdot sin(y)$ and $x \cdot cos(y)$ appear many times in this kind of symbolic mass matrices. If $x$ is a real-valued literal number very close to 0, with respect to an arbitrary absolute tolerance value \texttt{atol}, the entire node can be safely rewritten to 0 without causing substantial error propagation that invalidates the results of numerical evaluations, since $sin$ and $cos$ always return values between -1 and 1.
This optimization can then substantially reduce the size of the symbolic expressions contained in the matrix by propagating the simplification of multiplications where a factor has been rewritten to 0. In \autoref{fig:robotrules} we show the definition of the rewrite rules used by our custom symbolic optimizer. After computing all the equivalent symbolic expressions, our optimizer extracts the most promising expressions into a new, optimized symbolic matrix. Thanks to Julia's excellent compiler inspection tools provided in the standard library, the cost function (see \autoref{ssec:costfun} for details) we adopted for extraction approximates the cost of operations in a platform-dependent way by automatically counting how many assembly instructions a given mathematical operation compiles to.

\begin{figure}[htp!]
\begin{jllisting}[language=julia, style=jlcodestyle]
function optimizer_rules(atol=1e-13)
    @theory a b c begin 
        a + (b + c) == (a + b) + c
        a * (b * c) == (a * b) * c
        a * (b + c) == (a * b) + (a * c)
        a * 0 => 0
        a + 0 => a
        a::Real * cos(b) => 0 where isapprox(a, 0; atol=atol)
        a::Real * sin(b) => 0 where isapprox(a, 0; atol=atol)
    end
end
\end{jllisting}
\caption{Rewrite rules used by the custom optimizer.}
\label{fig:robotrules}
\end{figure}

We tested our optimizer in the same robot body dynamics system mentioned in \cite{gowda2021high}, considering only 2 degrees of freedom. In \autoref{tbl:robotbenchmarks} we show benchmarks of the numerical evaluation of the optimized symbolic mass matrix by varying the absolute tolerance value of the partial evaluator. The unoptimized computation required a mean evaluation time of 3.459 ms. We ran this simulation on a laptop running the Void Linux operating system \cite{voidlinux} with 16 Gigabytes of RAM and an 11th Generation Intel Core i7-1165G7 CPU running at a clock speed of 2.80GHz. On this machine, when generating a symbolic mass matrix of the system by choosing a higher number of degrees of freedom, the Julia compiler was not even able to compile an executable function to be evaluated numerically. When compiling, the machine ran out of available memory since the size of the symbolic terms in the unoptimized matrix easily exceeded several hundred Megabytes.

The first thing that can be observed from \autoref{tbl:robotbenchmarks}, is that by setting a larger absolute tolerance value, the optimizer prunes more branches of the symbolic expressions. This results in faster numerical evaluation but intuitively, larger error between the optimized and unoptimized computations.
Another thing that can be noticed is that for lower values of the absolute tolerance parameter, the average error stabilizes around $1.3\cdot 10^{-16}$. This additional error is most likely caused by the re-arrangement of the expressions caused by the commutativity, associativity and distributivity rewrites during the e-graph rewriting optimization phase.

\begin{table}[]
\begin{tabular}{|l|l|l|l|l|}
\hline
\begin{tabular}[c]{@{}l@{}}Absolute\\ Tolerance\\ Treshold\\ For Optimizer\end{tabular} & Samples & Average Error           & \begin{tabular}[c]{@{}l@{}}Mean Time\\ (Optimized)\end{tabular} & Mean speedup \\ \hline
$10^{-11}$                                                                              & 1241    & $3.752 \cdot 10^{-12}$  & 418.572 $\mu$s                                                  & 8.475x       \\ \hline
$10^{-13}$                                                                              & 923     & $4.710 \cdot 10^{-14}$  & 1.877 ms                                                        & 1.88x        \\ \hline
$10^{-15}$                                                                              & 862     & $4.319 \cdot 10^{-16}$  & 2.171 ms                                                        & 1.627x       \\ \hline
$10^{-18}$                                                                              & 831     & $1.355 \cdot 10^{-16}$  & 2.425 ms                                                        & 1.459x       \\ \hline
$10^{-20}$                                                                              & 853     & $1.337 \cdot 10^{-16} $ & 2.359 ms                                                        & 1.467x       \\ \hline
\end{tabular}
\caption{Numerical evaluation benchmarks of the symbolic mass matrix of a rigid body dynamics system with 2 degrees of freedom, optimized with our method. The mean evaluation time of the unoptimized matrix was 3.459 ms.}
\label{tbl:robotbenchmarks}
\end{table}

\FloatBarrier
\subsection{Chemical Reaction Networks with Catalyst.jl}
\label{ssec:catalyst}


In our paper "High-performance symbolic-numerics via multiple dispatch" \cite{gowda2021high}, first published online in May 2021, we reported how we used our proposed e-graph-based optimization method for Symbolics.jl expressions to test the effectiveness of these techniques in a real-world scenario, using the BioNetGen format to read in a 1122 ODE model of B Cell Antigen Signaling \cite{barua2012computational} and simplifying the 24388 terms using our method. The terms in the right-hand side contained only real-valued linear combinations, therefore, we managed to accelerate the generated code for the right-hand side from 15.023 $\mu$s to 7.372 $\mu$s per execution after a pass of e-graph optimization, effectively halving the time required to solve the highly stiff ODEs with ModelingToolkit.jl. In that experiment we used the Julia language version 1.5.3 and LLVM version 10. At the time of writing, the current stable Julia version is 1.6.3. 

\begin{table}[ht]
\begin{tabular}{|l|l|l|}
\hline
\begin{tabular}[c]{@{}l@{}}Average number of\\ LLVM IR instructions\end{tabular} & \begin{tabular}[c]{@{}l@{}}Not optimized with\\ Eq. Saturation\end{tabular} & \begin{tabular}[c]{@{}l@{}}Optimized with\\ Eq. Saturation\end{tabular} \\ \hline
LLVM Optimizations Disabled                                                      & 7228.1                                                                      & 4983.8                                                                  \\ \hline
LLVM Optimizations Enabled                                                       & 4238.5                                                                      & 4227.9                                                                  \\ \hline
\end{tabular}
\caption{Average number of LLVM IR instructions of code generated by fuzzed symbolic expressions with trivial operations, using Julia 1.6.1 and LLVM 11}
\label{tab:llvmlol}
\end{table}

Since we obtained this result in early 2021, newer versions of Julia and LLVM have caught up in terms of the quality of compiler optimizations.
We verified this by generating 100 random symbolic expressions with an AST depth of 12 levels, containing only real-valued symbolic variables from a set of 13 symbols, random constant floating point numbers and composite $f$-application terms with operations restricted to addition, multiplication and subtraction.
We then generated the Julia code for those expressions, both unoptimized and optimized with our equality saturation method. We lowered the generated Julia code to the LLVM IR, with optimizations passes both disabled and enabled.
In \autoref{tab:llvmlol} we show the average number of instructions obtained by lowering the code of the 100 randomly generated symbolic expressions to the LLVM IR, using Julia version 1.6.0 and LLVM version 11. Since the operations used in the randomly generated expressions are always lowered to a limited scope of trivial floating point instructions, not prone to error propagation, it is easy to conclude from the results in \autoref{tab:llvmlol} that for this task, optimizations in the Julia compiler pipeline that have been enabled in version 1.6 are achieving results as good as the ones we obtained with our method and Julia 1.5.




\chapter{Extensions and Future Work}
\label{ch:future}

In this chapter we will briefly describe possible extensions to the Metatheory.jl  package \cite{Cheli2021} that we introduced and discussed in \autoref{ch:solution}. We will then describe some possible future real-world applications of this expression rewriting and optimization framework. General-purpose term rewriting by equality saturation has been introduced very recently \cite{willsey2021egg} and is a technique with very interesting computational properties, leaving behind many open theoretical questions. Thus, we can go far beyond implementing compiler optimizers with this framework.
The context of the Julia programming language will allow researchers in the future to study and implement novel systems based on classical and e-graph  rewriting in a high-level fashion, without having to run into the annoying two-language problem. The adoption of Julia will also permit the study of those techniques in the presence of an expressive type system, powerful metaprogramming, code generation and performant numerical computing features.

\section{Metatheory.jl Improvements}

\subsection{Relational E-Matching}
An improvement to the e-graph pattern matching procedure (\autoref{ssec:ematching}) that has not yet been implemented in Metatheory.jl, has been proposed in \cite{zhang2021relational}. In their proposed technique, e-graphs are viewed as relational databases, effictively improving the overall time complexity and efficiency of the algorithm. Precisely, in the virtual machine e-graph pattern matcher \cite{de2007efficient}, equality checks are computed suboptimally. This new approach converts patterns into conjunctive queries, to be solved using a worst-case optimal join algorithm \cite{ngo2018worst} that deals with this flaw.

\subsection{Proof Production Algorithms}
\label{ssec:proofproduction}
Given a set of rewrite rules equality saturation can efficiently discover if two given symbolic expressions are equal. Equality saturations does this by trying to apply all possible rewrite rules on all the expressions contained inside of an e-graph. When two different expressions are initially inserted in an e-graph, they reside in two different e-classes. If after some iterations of equality saturation they end up resolving to the same e-class (see \autoref{defn:egraph}), it can be concluded that the two symbolic expressions are equal in respect to the given theory of rewrite rules. In its current state, though, the equality saturation implementation in Metatheory.jl is not able to produce an explanation \textit{a posteriori} of why two expressions became equal in the e-graph, since during every iteration, equality saturation tried to apply all the possible rewrites.
When Metatheory.jl is used to prove the equivalence of expressions in a domain-specific theory, having an algorithm that produces human-readable and machine-verifiable proofs would be of practical usage.  
Those proofs of equality can take the form of chains of directed rewrite rule applications: proofs in this format could be verifiable in polynomial time by using classical rewriting, while the equality saturation algorithm usually requires exponential time to discover the equality of two expressions.
Given a system of rules there could be many, possibly infinite ways of rewriting an expression into another. If equality saturation is able to find more than one,  a potential proof-production algorithm should extract the shortest proofs out of the e-graph. Proofs of equality can be achieved by keeping track of when and where rules have been applied in the e-graph, with respect to the congruence closure invariant maintenance steps. In fact, an entire trace of rule applications on an e-graph could in theory already be used as a machine verifiable proof, but there are substantial flaws in using those traces as proof certificates: they would not be human readable at all  and the time required to verify them would be approximately close to the time that equality saturation required to produce the traces.
There are already various proof-producing congruence closure algorithms in literature, such as the one proposed in \cite{nieuwenhuis2005proof}. The egg  \cite{willsey2021egg} developers are currently implementing such an algorithm in their e-graph term rewriting system, and a similar implementation based on  existing algorithms could be developed for Metatheory.jl in the near future.

\subsection{Smart Rule Scheduling Heuristics}
\label{ssec:futureschedulers}

The equality saturation framework is often referred to be an algorithm that involves an inefficient brute-force tactic of searching and applying matches in an e-graph. The main heuristic mechanism to prune the search space in terms of applicable rules (and therefore to avoid combinatorial explosion of e-graphs) is called \textit{rule scheduling}, already discussed in \autoref{ssec:schedulers}. 
A \textit{rule scheduler} is an arbitrary policy that at a given iteration of equality saturation, is responsible of disabling and enabling the search and application of certain rules in the system, basing this decision on arbitrary properties that can be inferred from the previous iterations. When feeding equality saturation with very large expressions and systems of rewrite rules, the algorithm can waste a lot of time in searching and applying rewrites that are practically useless for the end goal of the user.
The default and best available general purpose rule scheduler for Metatheory.jl uses an exponential backoff strategy that simply bans for a few iterations the rules that produced too many matches. This existing heuristic greatly improves the overall rewriting performance, but it is not informed at all about the users' goals and could thus perform much better by furtherly reducing the search space. Rule schedulers can be domain-specific in order to improve the performance of rewriting in certain theories, but the development of general rule schedulers can lead to huge performance improvements in the overall usage of Metatheory.jl.
Our goal is to develop \textit{goal-informed rule schedulers} that use specific heuristics to make equality saturation avoid rewrites that can be troublesome and most importantly, do not bring e-graphs in a state that is closer to users' end goals. 
A possible, general-purpose improvement to the exponential backoff policy could be informing the scheduler about the cost function that the user will later use for extracting expressions from an e-graph. Such a scheduler could avoid the repeated application of rules that did not previously reduce the overall value of the cost function of e-classes. When the user's goal is to prove the equivalence of two or more expressions, another potential policy could resort to an increased tendency of applying the rules whose patterns structurally resemble the expressions that the user wants to be prove equal, or to apply these rules only on related e-classes.
Later on, more modern artificial intelligence techniques can be adopted to improve goal-informed rule schedulers. Reinforcement learning \cite{sutton2018reinforcement} is a recently developed paradigm of machine learning concerned with how intelligent agents should take actions in an environment in order to maximize a reward. The rule scheduling problem could be formulated as a reinforcement learning problem where the rule scheduler (the agent) in a given e-graph and its history (the environment) has to decide at a given iteration if a rule should be applied or not (the possible actions). Game mastering reinforcement learning techniques such as AlphaZero \cite{silver2017mastering} or MuZero \cite{schrittwieser2020mastering} could be adopted by viewing the rule scheduling problem as a game that the agent can win, especially in an automated theorem proving context. Viewing e-graph rule scheduling as a reinforcement learning task is thus a very interesting open research problem that, together with a proof production algorithm (\autoref{ssec:proofproduction}), may yield groundbreaking results in many applications such as automated theorem proving (\autoref{ssec:theoremproving}), compiler optimizations and symbolic mathematics whereas equality saturation with classical heuristics may take a very long time to optimize and prove equalities in the presence of very large input expressions and systems of equational rewrite rules. 
This exciting potential application can be practically developed without running into the two-language problem: since Julia elegantly supports hardware accelerators such as GPUs \cite{besard2018effective} and is a language designed with high-performance numerical computation in mind, there are many excellent frameworks for machine learning and deep neural networks such as Flux.jl \cite{Flux.jl-2018, innes:2018}. Most importantly, there are already performant pure-Julia implementations of reinforcement learning algorithms such as AlphaZero \cite{Laurent2020AlphaZero}.

\subsection{Expression Language Grammars}
The egg library \cite{willsey2021egg} allows users to explicitly define the languages that form the symbolic expressions to be rewritten. This convenient feature is not only useful for statically checking the validity of rules and for preventing the creation of malformed terms while rewriting. 
When strict enough, explicit definition of an expression language helps by imposing a limit on the number of available function symbols that construct $f$-application terms, and most importantly imposes a constraint on their associated arity. Thus, all the operations and their associated arity in a language definition can be enumerated, and nested symbolic expressions can be efficiently represented in memory with flat arrays of positive integer numbers, instead of using the classical syntax-tree representation.  In performance-critical applications of equality saturation and e-graph rewriting, this internal representation of symbolic terms could provide substantial performance improvements.

\subsection{Automatic Parameter Inference}

The time required by equality saturation to saturate an e-graph is not easy to predict. 
Given an input set of rewrite rules, it may take milliseconds to saturate an e-graph from a starting expression, while by slightly changing the input expression or the set of rules, the e-graph may not even saturate at all.
Better understanding of when e-graph rewriting terminates by saturating is still an open research problem, but the fact that e-graph rewriting is a Turing-complete computational model (\autoref{ssec:turingcomplete}) strongly suggests that there will never be no algorithm for deciding when a given input expression and set of rules will cause an e-graph to saturate. Using a Turing-complete set of rewrite rules implies that the problem of deciding if an e-graph will saturate or not can be reduced to the halting problem \cite{turing1936computable}. More precisely, characterizing the situations when e-graph rewriting results in a combinatorial explosion for a "reasonable" rule-set, or more generally  characterizing the differences between e-graph and term rewriting, is also an unsolved research problem.
There have been discussions about why reasonable rule sets can create infinite loops in e-graph rewriting, and some properties of offending rule systems have been found\footnote{"Why reasonable rules can create infinite loops", at Github Discussions: \href{https://github.com/egraphs-good/egg/discussions/60}{https://github.com/egraphs-good/egg/discussions/60}}.
In real-world applications, such as symbolic mathematics or code optimization for improving simulations' performance, the speed of e-graph rewriting is important. This is why better rule scheduling heuristics (\autoref{ssec:schedulers}, \autoref{ssec:futureschedulers}) are also an important future improvement of Metatheory.jl. In those real-world applications, the size of input expressions and rule systems can be very large, and in the presence of associativity and commutativity. This can result in unpredictable behavior in the time required to rewrite an input expression into a goal form, and it will be very hard to infer if and how long it will  take for an e-graph to saturate. Since equality saturation may not terminate and the e-graph may blow up in a combinatorial explosion, users can provide size, iteration and time limits to eagerly halt equality saturation (\autoref{fig:saturationparams}), and since the behavior of an e-graph is strictly dependent on the rule set and input expressions, some parameters may provide good performance for some input expressions, while resulting in poor performance for some other expressions by rewriting with the same rule set. 
This is obviously not convenient, since users may have to manually change the parameters to avoid halting equality saturation too early (and thus not reaching their rewriting goal), or letting it run for too long, effectively consuming a lot of computational resources and wasting precious time.
The open question is to verify if any automatic parameter inference technique can be repurposed to automatically infer the best default parameters for equality saturation and schedulers in specific contexts, given the set of rewrite rules that will be applied and optionally an input language grammar.

\section{Theoretical Applications}

\subsection{E-Graph Rewriting is Turing-Complete}
\label{ssec:turingcomplete}
Intuitively, given an appropriate collection of rewrite rules, e-graph rewriting is Turing-complete just as classical term rewriting is. To show this with Metatheory.jl, we have implemented interpreters for a minimal Turing-complete programming language, called the WHILE language, introduced by Professor Pierpaolo Degano in his computability theory lecture notes \cite{deganoecc}. 
We defined two interpreters that use rewrite rules based on the denotational semantics of this language, one interpreter relying on classical rewriting\footnote{Source code for the classical interpreter available at \href{https://git.io/JXbmm}{https://git.io/JXbmm}} and one relying on e-graph rewriting \footnote{Source code for the nondeterministic interpreter available at \href{https://git.io/JXbmC}{https://git.io/JXbmC}}. The interesting property that we have observed is that the implementation based on e-graph rewriting allows for \textit{nondeterministic} and \textit{reversible} computation, with an obvious exponential tradeoff in time and space complexity when compared to the classical rewriting interpreter. Another interesting observation we made, is that this e-graph rewriting interpreter allows for, other than retrieving the result of the evaluation of a program, to retrieve all the intermediate states that the program has gone through the computation. This being due to the property of e-graphs of being \textit{monotonic}, (in short, never destroying the information stored when rewrites are applied).
Together with the addition of a proof-production algorithm (\autoref{ssec:proofproduction}) and extensible rule scheduling heuristics (\autoref{ssec:futureschedulers}), e-graphs are a promising framework for modeling algorithms and computational models designed to solve problems in NP and other complexity classes that involve nondeterministic computation. 
Equality saturation is thus a general-purpose technique that can model the execution of nondeterministic computations, requiring exponential time in order to simulate the execution of nondeterministic term-rewriting systems. 
A proof-production algorithm (\autoref{ssec:proofproduction}) for e-graph rewriting would produce a list of the directed rewrite rule applications required to transform an expression into a resulting equivalent expression. These 'proofs', made of chains of rewrites are executable in polynomial time in a classical rewriting context. When using e-graph rewriting as a framework to model algorithms that solve NP problems, these automatically produced rewrite proofs could play the role of certificates verifiable in deterministic polynomial time.
Rule schedulers (\autoref{ssec:futureschedulers}) can then be a very high-level interactive abstraction to formulate efficient heuristics to solve NP problems.
On the whole, this framework may not be the most efficient way of simulating nondeterministic computations, but the very high-level abstractions required to describe such computations can help theoretical computer scientists to substantially reduce the time between a first formulation of computational problems and the actual simulation of algorithms. Thus, the built-in proof production and rule scheduling heuristics are given "for-free" with e-graph rewriting, and this overall framework may aid theoretical computer scientists in discovering novel and more efficient algorithms to solve problems in NP. 

\subsection{Algebraic Metarewriting}
Rewrite rules and equations are themselves symbolic expressions. In fact, the concrete types that represent rewrite rules in Metatheory.jl implement the methods for TermInterface.jl (\autoref{sec:terminterface}). Therefore, rewrite rules in our system can be directly used as symbolic expressions to be rewritten. This opens up a world of opportunities that can be named \textit{algebraic metarewriting}. If metaprogramming is the practice of writing computer programs that manipulate other computer programs, \textit{metarewriting} is the practice of writing term rewriting systems that manipulate other rewriting systems. The equational nature of our framework is particularly adapt for prefixing the adjective \textit{"algebraic"} to the name of these practices. 
Authors of egg \cite{willsey2021egg} have recently published a paper introducing a technique that is using equality saturation to automatically infer new systems of rewrite rules \cite{nandi2021rewrite}. Rewrite rules and systems can be treated as algebraic entities much like terms in symbolic mathematics. Combining this potential application of Metatheory.jl with other use cases and improvements such as automated theorem proving and reinforcement learning scheduling can open a pandora's box, a world of opportunities, so vast that at least in this thesis, deserves its own name.


\FloatBarrier

\subsection{Programming Language Theory and Julia}
\label{ssec:plt}

The Julia programming language is a general-purpose language that has been used for innumerable scientific applications. Due to the lack of some built-in language features such as pure algebraic data types and pattern matching, the Julia language hasn't attracted many users from the field of PLT (Programming Language Theory). The development of interpreters and compilers in the Julia language can be impractical and tedious to develop without high-level language constructs for pattern matching, especially for users that are not familiar with the paradigm of multiple dispatch. 
Compilation, abstract and concrete interpretation tasks can still be elegantly implemented in Julia's flexible type system and multiple dispatch paradigm.
Metaprogramming in Julia though, opens up some opportunities that can save a lot of precious programming language theorists' time. There is no need of designing a completely new programming language every time: a Julia eDSL with a customized interpreter, overriding Julia's internal semantics, can often suffice and may help save a lot of time compared to implementing experimental programming languages with novel constructs from scratch. Most importantly, implementing PLT experiments within Julia eDSLs can spare researchers from the tedious task of having to implement a new parser and to replicate or wrap all the built-in functionalities that are available in a programming language standard library. To practitioners that do not want to adopt Julia's syntax for their programming languages, there are various Julia packages available, offering lexing utilities and parser combinators \footnote{An example of a package offering modern lexing and parsing functionalities is RBNF.jl, available at \href{https://github.com/thautwarm/RBNF.jl}{https://github.com/thautwarm/RBNF.jl}}.

Thankfully, Julia is extremely extensible through metaprogramming and the macro system, and there are many available Julia packages providing eDSLs (embedded domain specific languages) that are designed for a variety of purposes, such as introducing pattern matching constructs in Julia.
The first-class pattern matching features proposed in our system can be used to implement programming language interpreters in a surprisingly elegant way. See for example the source code for the experimental interpreters we implemented with Metatheory.jl, mentioned in \autoref{ssec:turingcomplete}: the syntax of rewrite rules that define the interpreter closely resembles the denotational semantics on which the language is based \cite{deganoecc}. 
Metatheory.jl's rule definition language and functional combinators actually generalize and extend the features that can be found in languages with built-in pattern matching like in the ML family. The behaviour of such pattern matching constructs can be therefore easily recreated with our system by using a \texttt{Chain} functional rewriter combinator (\autoref{ssec:combinators}) together with a system of only \texttt{DynamicRule}s (\autoref{ssec:patternsrules}), but intuitively, the expressive power of all the other functional combinators and rewrite rule types is greater. Also, Metatheory.jl allows to unleash the power of e-graph rewriting in the context of programming language theory and practice.

\section{Practical Applications}

\subsection{Automated Theorem Proving in Pure Julia}
\label{ssec:theoremproving}

E-graphs have been extensively used \cite{de2007efficient} in ATPs (Automated Theorem Provers) environments such as Z3 \cite{de2008z3} and Simplify \cite{detlefs2005simplify}. Those solvers
adopt a technique called DPLL(T) (DPLL modulo theories \cite{nieuwenhuis2005abstract}) to solve the SMT (Satisfiability modulo theories \cite{barrett2018satisfiability}) problem, a generalization of the classical boolean satisfiability (SAT) problem in the presence of domain-specific theories. DPLL(T) stands for an extension of Davis–Putnam–Logemann–Loveland (DPLL) algorithm \cite{davis1960computing} for deciding the satisfiability of propositional logic formulae in conjunctive normal form, but in the presence of a domain-specific theory. After more than 50 years, the DPLL procedure is still extensively used by many efficient and complete SAT solvers. In this extension, the DPLL(T) solvers repeatedly consult domain-specific theory solvers to
check the consistency of queries under a domain-specific theory, often specified as input by users of the ATP environments.
Z3 \cite{de2008z3} and other ATPs are often used in software verification tasks. A common technique is to translate assertions about code into SMT formulae in order to determine if all properties can hold.   
A pure Julia ATP that emphasizes on metaprogramming could help solving the two-language problem in the context of program verification, providing a highly generic framework for static analysis of Julia code. A fundamental strength of the program optimization framework we describe in this thesis is that users can define high-level domain-specific compiler optimizations directly inside the same programs they are writing. A pure Julia ATP that supports homoiconic Julia expressions could be used to achieve program verification with a similar paradigm. Programmers could define domain-specific constraints in the same program they want to verify, and then define a macro to use in critical sections of the program that can preemptively halt compilation if the user-defined constraints and assertions cannot be solved.
The generalized equality saturation rewriting framework can be used to practically solve ground-term unification problems \cite{martelli1982efficient} and to prove the equivalence of symbolic expressions. 
A simple proof-production algorithm (\autoref{ssec:proofproduction}) could then produce human-readable and machine-verifiable proofs in the form of chains of directed rewrites that are necessary to make two or more symbolic expressions equal in a given domain-specific theory.
We have experimented with very simple deductive theorem proving tasks in propositional logic modulo theory \footnote{Source code of the experiment is available at \href{https://git.io/JXbXU}{https://git.io/JXbXU}} by using homoiconic Julia expressions. Our equality saturation framework provides great generality but is obviously not the most performant solution to solve the SMT problem. However, for numerous applications in software verification and analysis, proving only propositional ground formulae is insufficient as extensive quantifier reasoning is often needed. 
Theorem provers like Simplify \cite{detlefs2005simplify} that adopt a DPLL(T) strategy, use e-graphs and the e-matching algorithm (\autoref{ssec:ematching}) to instantiate quantified variables when integrating domain-specific theory reasoning. These techniques have been used in practice for a long time before the very recent generalization of equality saturation on e-graphs has been introduced for general-purpose term rewriting. 
The generic implementation of e-graphs and equality saturation in Metatheory.jl could be extended to implement a flexible, pure-Julia, highly generic ATP environment that proves formulae in fractions of first (or even higher) order logics by first checking consistency in domain-specific equational theories and then producing conjunctive queries when needed, to be solved by classical SAT solving algorithms that can perform more efficiently than a pure term rewriting solutions.
A pure Julia ATP would not only be useful for program verification. Such a system could also be used to perform constraint solving in a symbolic mathematics context. Such as the simplification engine in the Symbolics.jl \cite{gowda2021high} (\autoref{sec:symbolics}) computer algebra system. There have been experiments in extending Symbolics.jl to support constraint solving through the Z3 theorem prover \footnote{SymbolicsSAT.jl extends Symbolics.jl simplification with a theorem prover \href{https://github.com/JuliaSymbolics/SymbolicSAT.jl}{https://github.com/JuliaSymbolics/SymbolicSAT.jl}}, but this approach suffers from the two language problem: it requires sofisticated conversion of symbolic Julia expressions into Z3 queries and would not benefit at all from the symbolics-metaprogramming synergies of the framework we proposed in this thesis,

\subsection{Automatic Floating-Point Error Fixers}

Scientific computing depends on floating point arithmetic to approximate real valued arithmetic. This implicit approximation introduces rounding error that can accumulate and propagate to produce unacceptable results. The literature in  numerical methods provides many techniques to mitigate this error propagation, but applying these techniques manually requires programmers to understanding details of floating point arithmetic and to rearrange expressions accordingly. 
Herbie \cite{panchekha2015automatically} is a tool that uses equality saturation and advanced heuristics to automatically rewrite mathematical expressions to improve their floating-point accuracy. The core of Herbie has been recently rewritten to use \cite{willsey2021egg}. Herbie is written in a mixture of the Racket, Scheme and Rust programming languages, and accepts input expressions in a specific Scheme-like format. This requires programmers that want to use this technique in their programs to install the language runtimes, install the Herbie package, convert the symbolic mathematical expressions into this Scheme-like format, execute the Herbie rewriter and then convert back the results into their starting format. This can be highly unpractical for non-experts. This technique could be theoretically ported to a pure Julia solution that would harmonically coexist with the built-in language metaprogramming system. Just like our example stream fusion optimizer in \autoref{sec:streamfusion}, the exposed interface of this floating-point error fixing system in Julia could consist of a single macro that takes a Julia expression and substitutes it with a fixed one, just before the piece of code is actually compiled and executed.

\subsection{Simplification Engine for Fock Algebras}
\label{ssec:quantum}
The Julia package QuantumCumulants.jl \cite{plankensteiner2021quantumcumulants} provides symbolic derivation utilities for mean-field equations for quantum mechanical operators in Julia. The algebraic framework for modeling open quantum systems provided by QuantumCumulants.jl is an extension of the Symbolics.jl \cite{gowda2021high} (\autoref{sec:symbolics}) Julia CAS, providing an implementation of specialized symbolic expression types and classical rewriting simplification rules for the canonicalization of expressions in the specialized  non-commutative algebras that are required to describe quantum systems mathematically. 

This software automatically derives mean-field equations from the symbolic description of quantum systems. Those equations are automatically expanded in terms of cumulants to an arbitrary order, resulting in a closed set of symbolic differential equations that can be solved numerically by ModelingToolkit.jl \cite{ma2021modelingtoolkit} (\autoref{sec:mtk}) and DifferentialEquations.jl \cite{rackauckas2017differentialequations}. The general picture of the workflow is analogous to how users typically model other kinds of systems with Symbolics.jl and ModelingToolkit.jl.
Users give a high-level symbolic description of the system, and a system of differential equations is automatically derived or explicitly given. Such collection of differential equations is then automatically transformed and  compiled to fast, executable Julia code that evaluates a time step of the system numerically. The automatically generated code can then be fed into state-of-the-art advanced numerical differential equation solvers \cite{rackauckas2017differentialequations}.
Thus, our automatic domain-specific code optimization solution described in \autoref{ch:solution}, suitable for both Julia code and Symbolics.jl expressions, can be intutively adopted for QuantumCumulants.jl for both preemptive optimization, simplification and expansion of symbolic expressions before the actual numerical simulations, with an overall approach similar to the one we described in \autoref{sec:mtk} and \autoref{ssec:catalyst}. 
An exciting future application of our methodology could consist in the development of a specialized set of equational rewrite rules for QuantumCumulants.jl, together with a composable set of symbolic simplification combinators, with the end goal of improving both the accuracy and the execution time of the numerical simulations of open quantum systems. This would likely be the first implementation (that we are aware of in literature) of a symbolic simplification engine for Fock algebras.
The same concepts could practically apply to other packages that provide extensions of Symbolics.jl to model systems in various scientific domains of study.

\subsection{Other Optimization Applications}
\label{ssec:otheroptimizers}

In \cite{yang2021equality} authors propose a superoptimizer for deep learning tensor graphs, built on top of the egg \cite{willsey2021egg} library for equality saturation rewriting. In \cite{wang20spores} authors introduce a general optimization technique for linear algebra expressions by still relying on egg. 
Julia has excellent built-in support for linear algebra, n-dimensional tensors, hardware accelerators and numerical computation primitives. All these equality saturation techniques for code optimization could be ported to Metatheory.jl and adapted to a pure Julia context. Thanks to the metaprogramming system and homoiconicity, each one of these optimizers would harmonize well with the others.
Additionally, Julia programmers are able to reflectively inspect and manipulate the lowered intermediate representations of code before it gets compiled. In the original paper introducing equality saturation \cite{tate2009equality} authors adopted an intermediate representation of programs called Program Expression Graphs (PEGs) to encode loops and conditionals, effectively encoding the Control Flow Graph (CFG) of a program in a purely functional form, a format that plays well with e-graphs and the equality saturation workflow. Such a PEG representation could be implemented directly for lowered Julia IRs, with the aim of allowing the Julia compiler to execute plug-and-play extensible code optimizers, obviating the need to worry about optimization ordering.

\chapter{Conclusion and Acknowledgments}
\label{ch:conclusion}
\section{Evaluation and Final Remarks}

In this thesis, we have presented a novel system for advanced expression rewriting in the Julia programming language. We have presented a flexible generic interface for symbolic expressions in \autoref{sec:terminterface}, and we have presented our framework in \autoref{sec:metatheory}, designed to generalize classical rewriting with a powerful functional abstraction, and to extend the Julia language with an extensible rewrite rule definition language that shares many features with pattern matchers in other functional programming languages, thus introducing a new language construct in the host Julia language, thanks to its prodigious extensibility via the Scheme-like macro system. 
The main component of our system is an extended general-purpose implementation of the novel technique of term rewriting via equality saturation (\autoref{ssec:eqsat}), an algorithm that uses the e-graph (\autoref{ssec:egraphs}) data structure to conveniently store and rewrite on many equivalent symbolic expressions in a nondeterministic fashion, providing a robust layer of abstraction over term rewriting that allows to unleash the full expressive power of algebraic descriptions. Thus, users can provide definitions of rewrite systems by writing mathematical identities, the same way in which us humans understand, read and write mathematics. 
This e-graph rewriting technique obviates the need to worry about rewrite ordering, or more precisely, having to convert mathematical theories to directed, confluent and terminating rewriting systems, in order to implement symbolic expression manipulation in the context of domain-specific algebras. 
One of our goals was to solve the two-language problem in the context of domain-specific compiler optimization: many other tools that rely on equality saturation to rewrite programs such as Herbie \cite{panchekha2015automatically} or TENSAT \cite{yang2021equality}, often resort to implementing compiler optimizations in many languages, relying on the Rust egg \cite{willsey2021egg} library for general purpose equality saturation. Those systems manipulate terms in an S-expression format, while implementing the other principal components in Lisp-like languages suitable for classical term rewriting and compiler development, such as the Racket dialect of Scheme. Those program rewriters can often only rewrite programs written in special Lisp-like DSLs. To actually optimize code in real-world scientific computing applications, those systems require extensive transpiling infrastructure in order to convert between S-expressions and programs in different target languages, such as Python. 
The consequences of the two-language problem in the context of domain-specific program rewriting are directly observable: these program rewriting systems are highly complex, hardly extensible and made of many intercommunicating modules and wrappers written in different programming languages. They are thus hard to debug and to read, and require great expertise to be correctly designed and developed. 
This instance of the two-language problem suppresses generality, slows down development and experimentation, can cause architectural instabilities and hidden bugs, and most importantly requires the users to be programming language experts in order to apply the equality saturation code rewriting technique to their domain of study. 
We have shown how our framework practically solves the two-language problem in advanced program rewriting, analysis and optimization. The core implementation of Metatheory.jl is less than 4000 lines of thoroughly documented, pure Julia code, written to be as readable and extensible as possible by users and Open-Source contributors.

Other than solving the two-language problem for domain-specific compiler optimizations, our proposed expression rewriting framework practically demonstrates that by relying on a programming language with enough expressive power, it is possible to adopt a generic interface to bridge to provide the gap between compiler optimizations and symbolic mathematics. Thanks to TermInterface.jl (\autoref{sec:terminterface}), we were able to redesign the whole architecture of the Julia Symbolics ecosystem of packages (\autoref{sec:architecture}), and we were able to port many generalizable features to our package Metatheory.jl, in order to provide strikingly generic, low-level abstraction for term-rewriting. The pattern matching system is able to match and rewrite on both homoiconic Julia expressions (\autoref{fig:mtexample}, \autoref{sec:streamfusion}) and specialized symbolic expressions from the Symbolics.jl \cite{gowda2021high} computer algebra system (\autoref{sec:symbolics}), all with the same rule definition language.

In \autoref{ch:applications} we demonstrated that our framework has many practical applications in the Julia programming language.
We have shown an example optimizer that rewrites streams in Julia programs written in a functional paradigm (\autoref{sec:streamfusion}), and we have shown how integrating and merging our techniques with the classical rewriting features of Symbolics.jl, has allowed us to create domain-specific compiler optimizations for state-of-the-art symbolic-numeric simulations, in the domain of robotics (\autoref{ssec:robot}) and systems biology (\autoref{ssec:catalyst}). 

To conclude, in \autoref{ch:future} we illustrate some flaws of our system and what can be some future improvements in order to improve the expressiveness power and performance of the core of Metatheory.jl. E-graph pattern matching is NP-hard, and we thus show how in the future we could implement a novel technique that views the problem as solving conjunctive queries on relational tables \cite{zhang2021relational}. We briefly described how the core e-graph rewriting system can be extended with a proof-production algorithm (\autoref{ssec:proofproduction}) that lets equality saturation search in the space of all possible rewrites with exponential time, to prove that two or more expressions are equal, and can then return a human-readable and machine-verifiable proof (in polynomial time), in the form of a chain of the rewrite rules that have to be applied in order to rewrite an expression $A$ to an expression $B$. This allows for further exploration of the immensely vast applications of e-graph rewriting, beyond the scope of compiler optimizations and towards novel methodologies in automated theorem proving. We then show what are some future applications of this flexible technique in the Julia programming language. Many other optimizers for domain-specific symbolic-numeric simulations could be implemented in the future (\autoref{ssec:quantum}, \autoref{ssec:otheroptimizers}), and many techniques that have been already implemented in other programming languages with egg could be ported to pure Julia, harmonizing with the other symbolic manipulation and compiler optimizations solutions implemented with our system.

\section{What I Learned}

During the period of time in which I wrote Metatheory.jl and this thesis, approximately the course of a year, I have learned innumerable new things, and I have been able to apply almost everything that was taught to me during my 3 years of undergraduate computer science studies at the University of Pisa. I am therefore obliged to thank everybody who helped me in this long path. I discovered the Julia programming language by accident, and during the last year I dived into learning this new language as much as I could. Learning Julia, an incredibly expressive and general language, and developing the Metatheory.jl system, helped me learn how to practically apply concepts covered by many of the courses I've attended: algorithms and data structures, computability theory, programming language theory, classical artificial intelligence, software engineering, calculus, linear algebra, discrete mathematics, logic and numerical computing. The things I've learned that weren't covered by my university's courses were the Julia programming language, advanced metaprogramming, term rewriting concepts, symbolic mathematics and how symbolic-numeric simulations work, also, I've learned the role of such simulations in solving real-world scientific problems.

\section{Acknowledgments}

First, I have to thank my family for loving me and having supported me in every way through my three years of undergraduate studies. \\

I then have to thank my advisors Gian-Luigi Ferrari, Christopher Rackauckas and Andrea Corradini. 
I then have to thank professors Gian-Luigi Ferrari and Pierpaolo Degano for teaching my favorite courses at university.
I then have to thank Peter Czaban and the other members of Planting Space for all the interest they have shown in this project, the opportunity to collaborate with them and for the financial support by sponsoring Metatheory.jl on Github.
An important acknowledgment has to go to Jeff Bezanson, Stefan Karpinski, Viral B. Shah and Alan Edelman for designing the Julia programming language, a project that stands out from the crowd. I also have to thank all the other 1204 Open-Source contributors to the Julia programming language. I'd love to continue contributing to Julia projects and to continue my studies and research with this wonderful programming language as a companion, a language that has allowed me to express my thoughts in an impressively direct way: a language not only made for taming machines, but an awesome language that can bring people together and create decentralized global communities of programmers and researchers, sharing the goal of communicating through very expressive code. 
Special thanks go to Philip Zucker for making me discover the original egg paper and for having the original idea of implementing e-graphs in Julia, without Philip sending a message in a group chat on the Julia Zulip community at the beginning of 2021, this project would never have existed. Philip and I developed the first experimental versions of Metatheory.jl together, and then held night-long chats about e-graphs and other marvelous computer science topics. 
I'd like to thank Max Willsey, Chandrakana Nandi, Yisu Remy Wang, Oliver Flatt, Zachary Tatlock and Pavel Panchekha for authoring egg, the first-ever implementation of equality saturation for general-purpose term rewriting. 
I have to thank Philip Zucker, David P. Sanders, Max Willsey and James Fairbanks for having reviewed my first paper I ever published: "Metatheory.jl: Fast and Elegant Algebraic Computation
in Julia with Extensible Equality Saturation" \cite{Cheli2021}. 
I have to thank Shashi Gowda and Christopher Rackauckas for all the interest they showed in my project and for first having the idea of using Metatheory.jl together with Symbolics.jl and ModelingToolkit.jl. Shashi and Chris first invited me to experiment with this integration, and then constantly guided me through understanding all the internals of the systems they designed at MIT, Julia Labs and Julia Computing. I have to thank Shashi and Chris for inviting me to be a member of the Julia Symbolics Open-Source organization, and I also have to thank them especially for inviting me to coauthor my second-ever paper with them, the canonical citation for the Symbolics.jl system: "High-performance symbolic-numerics via multiple dispatch" \cite{gowda2021high}. I also have to thank Yingbo Ma, Maja Gwozdz, Viral B. Shah and Alan Edelman for the amazing opportunity of coauthoring this paper with them. 
Special thanks go to my friend Filippo Bonchi. Filippo showed a lot of interest in the project, helped me with a friendly review of my first paper \cite{Cheli2021} and taught me a lot of things about logic, category theory and how to author scientific papers.
I surely have to thank Chen Zhao, Philip Zucker, David P. Sanders,  McCoy R. Becker,  Jesse Perla, 'NumHack', 'fengmingze', Mosè Giordano, Greg Peairs, Jiayi Wei and Shashi Gowda for personally contributing to the development of Metatheory.jl. I also have to thank all the members of the Julia Symbolics and SciML organizations and all the Open-Source contributors that are developing the packages mentioned in this work, where Metatheory.jl is integrated and used, such as Symbolics.jl, SymbolicUtils.jl, ModelingToolkit.jl and DifferentialEquations.jl and all the other software works. 
I then have to personally thank my friends Ruben, Dario, Sabin, Gaia, Roberta, Andrea, Bianca, Valerio, Gabriele, Raffaele, Rossella, Cristina, Alberto, Michi, Caligula, Davide, Federica, Alessio, Giovanna, Piero, Daniele, Matilde, Ilaria, Enrico, Antonella, Adele, Fabio, Alessandro, Gabriele, Andrea, Giovanni, Pietro, Rachele, Francesca, Francesco, Eugenio, Nyaso and Rei.

\printbibliography

\end{document}